\documentclass[twocolumn,trackchanges]{aastex62}
\usepackage{bm}
\usepackage[colorinlistoftodos]{todonotes}
\hypersetup{colorlinks,citecolor=blue,linkcolor=red}
\usepackage{amsmath}

\revised{\today}
\accepted{October 30, 2019}
\shorttitle{neutrino nucleon-nucleon bremsstrahlung}
\shortauthors{Guo and Mart\'inez-Pinedo}
\begin{document}

\title{Chiral Effective Field Theory Description of Neutrino
  Nucleon-Nucleon Bremsstrahlung in Supernova Matter}

\correspondingauthor{Gang Guo}
\email{gangg23@gmail.com}%{g.guo@gsi.de}

\author[0000-0003-0859-3245]{Gang Guo}
\affiliation{GSI Helmholtzzentrum f\"ur Schwerionenforschung,
  Planckstra{\ss}e~1, 64291 Darmstadt, Germany}
 % \affiliation{Institutf{\"u}r Kernphysik (Theoriezentrum), Technische Universit{\"a}t Darmstadt, Schlossgartenstra{\ss}e 2, 64298 Darmstadt, Germany}

\author[0000-0002-3825-0131]{Gabriel Mart\'inez-Pinedo}
\affiliation{GSI Helmholtzzentrum f\"ur Schwerionenforschung,
  Planckstra{\ss}e~1, 64291 Darmstadt, Germany}
\affiliation{Institut
  f{\"u}r Kernphysik (Theoriezentrum), Technische Universit{\"a}t
  Darmstadt, Schlossgartenstra{\ss}e 2, 64298 Darmstadt, Germany}

%% Note that the \and command from previous versions of AASTeX is now
%% depreciated in this version as it is no longer necessary. AASTeX 
%% automatically takes care of all commas and "and"s between authors names.

%% AASTeX 6.2 has the new \collaboration and \nocollaboration commands to
%% provide the collaboration status of a group of authors. These commands 
%% can be used either before or after the list of corresponding authors. The
%% argument for \collaboration is the collaboration identifier. Authors are
%% encouraged to surround collaboration identifiers with ()s. The 
%% \nocollaboration command takes no argument and exists to indicate that
%% the nearby authors are not part of surrounding collaborations.

%% Mark off the abstract in the ``abstract'' environment. 
\begin{abstract}
  We revisit the rates of neutrino pair emission and absorption from
  nucleon-nucleon bremsstrahlung in supernova matter using the
  $T$-matrix formalism in the long-wavelength limit. Based on two-body potentials
  of chiral effective field theory ($\chi$EFT), we solve the
  Lippmann-Schwinger equation for the $T$-matrix including
  non-diagonal contributions. We consider final-state Pauli blocking
    and hence our calculations are valid for nucleons with an arbitrary
    degree of degeneracy. We also explore the in-medium effects on
  the $T$-matrix and find that they are relatively small for supernova
  matter. We compare our results with one-pion exchange rates,
  commonly used in supernova simulations, and calculations using an
  effective on-shell diagonal $T$-matrix from measured phase
  shifts. We estimate that multiple-scattering effects and
  correlations due to the random phase approximation
  introduce small corrections on top of the $T$-matrix
  results at subsaturation densities. A numerical table of the
    structure function is provided that can be used in supernova
    simulations. 
\end{abstract} 

%% Keywords should appear after the \end{abstract} command. 
%% See the online documentation for the full list of available subject
%% keywords and the rules for their use.


%% From the front matter, we move on to the body of the paper.
%% Sections are demarcated by \section and \subsection, respectively.
%% Observe the use of the LaTeX \label
%% command after the \subsection to give a symbolic KEY to the
%% subsection for cross-referencing in a \ref command.
%% You can use LaTeX's \ref and \label commands to keep track of
%% cross-references to sections, equations, tables, and figures.
%% That way, if you change the order of any elements, LaTeX will
%% automatically renumber them.
%%
%% We recommend that authors also use the natbib \citep
%% and \citet commands to identify citations.  The citations are
%% tied to the reference list via symbolic KEYs. The KEY corresponds
%% to the KEY in the \bibitem in the reference list below. 

\section{Introduction} \label{sec:intro}
    
Neutrino interaction with nucleons in proto-neutron stars
(PNS)~\citep{Burrows.Reddy.Thompson:2006} plays a crucial role in many
aspects of core-collapse supernovae (CCSNe), such as the explosion
mechanism~\citep{Burrows:2013,Janka.Langanke.ea:2007,Janka:2012} as
well as the synthesis of heavy elements in neutrino-driven
winds~\citep{Arcones.Thielemann:2013,Martinez-Pinedo.Fischer.ea:2016}
and the long-term cooling of the neutron star~\citep{Yakovlev.Kaminker.ea:2001,Yakovlev.Pethick:2004}. Three-dimentional
(3D) simulations with detailed neutrino transport have shown that
explosions are very sensitive to neutrino opacities even at the level
of 10\%--20\% \citep{Melson.Janka.ea:2015,Burrows.Vartanyan.ea:2018}. Therefore, an accurate
description of neutrino interaction in hot and dense nuclear matter
related to CCSNe is highly demanded.

We revisit the neutrino pair emission and absorption from NN collision
in supernova (SN) matter using $T$-matrix elements based on $\chi$EFT potentials
following~\citet{Bartl.Pethick.Schwenk:2014,Bartl:2016}. Neutrino
bremsstrahlung $NN\to NN\nu\bar\nu$, its inverse
$NN\nu\bar\nu \to NN$, and the related inelastic scattering
$NN\nu \to NN\nu$ play key roles in changing the number density and
energy for the heavy-flavor SN neutrinos and are thus important in
determining the neutrino spectra
formation~\citep{Raffelt:2001,Keil.Raffelt.Janka:2003}. The most
widely used bremsstrahlung rate in SN
simulations~\citep{Hannestad.Raffelt:1998} is based on the
one-pion exchange (OPE) potential in the Born approximation with only
interactions among neutrons considered. As already mentioned
by~\citet{Hannestad.Raffelt:1998}, a proper treatment of NN
correlations for general nuclear matter should be considered for a
better description of neutrino bremsstrahlung~\citep[see
also][]{Friman.Maxwell:1979,Sigl:1997,Yakovlev.Kaminker.ea:2001,Bartl.Pethick.Schwenk:2014,Pastore.Davesne.Navarro:2015,Dehghan.Moshfegh.Haensel:2016,Dehghan.Moshfegh.Haensel:2018,Riz.Pederiva.Gandolfi:2018}. Modern
nuclear interactions from $\chi$EFT have been used to study neutrino
bremsstrahlung based on the Landau's theory of Fermi
liquids~\citep{Lykasov.Pethick.Schwenk:2008,Bacca.Hally.ea:2009,Bacca.Hally.ea:2012,Bartl.Pethick.Schwenk:2014,Bartl:2016}. The
necessity to go beyond the Born approximation was demonstrated
by~\citet{Bartl.Pethick.Schwenk:2014} using effective on-shell
$T$-matrix elements extracted from experimental phase
shifts~\citep[see also][]{Sigl:1997,Hanhart.Phillips.Reddy:2001,Dalen.Dieperink.Tjon:2003}. It
should be pointed out~\citep{Bartl.Pethick.Schwenk:2014}, however, that the use of the on-shell $T$-matrix is
only valid in the limit of zero energy transfer between nucleons and
the neutrino pair. For finite energy transfer, off-shell $T$-matrix
elements are needed. \citet{Dalen.Dieperink.Tjon:2003} also explored
the in-medium effects on the $T$-matrix based on the Bonn C potential
for neutrino bremsstrahlung rates, but their study was limited to
neutrino emissivities in conditions relevant to neutron stars. \citet{Bartl.Pethick.Schwenk:2014} performed the
first calculation of NN bremsstrahlung for arbitrary mixtures of
neutrons and protons in supernova matter.

In this work we aim for an improved description of neutrino
bremsstrahlung that includes both off-shell matrix elements and Pauli
blocking effects. We solve the Lippmann-Schwinger (LS) equation to
obtain the vacuum $T$-matrix~\citep{Lippmann.Schwinger:1950} and the
Bethe-Goldstone (BG) equation~\citep{Bethe:1956,Goldstone:1957} to
account for in-medium effects in the $T$-matrix. The bremsstrahlung
rate, or more precisely the associated structure function
$S(q, \omega)$, with $q$ and $\omega$ the momentum and energy
transfer, is obtained using the Fermi's golden rule in the long
wavelength limit ($q\to 0$), which is consistent with that derived
from the finite-temperature linear response theory~\citep[see,
e.g.,][]{Weldon:1983,Roberts.Reddy:2017}. To account for
multiple-scattering effects and to get around of divergences at
$\omega\to0$, we introduce a relaxation rate parameter or width
parameter whose value is determined from the normalization of
$S(q\to 0, \omega)$~\citep{Hannestad.Raffelt:1998}. Our calculations consider final-state
  blocking for the nucleons in calculating the bremsstrahlung
  rates. They are compared to results using Boltzmann distributions
  without blocking, which are only valid in the non-degenerate
  regions.
          
The paper is organized as follows. In Sec.~\ref{sec:brem}, we
calculate perturbatively the structure function and the neutrino
bremsstrahlung rate, and then study the effects of using different
nuclear matrix elements (vacuum $T$-matrix, in-medium $T$-matrix and
OPE potential) with/without blocking, and with half-off-shell or
on-shell matrix elements. In Sec.~\ref{sec:norm}, we include the width
parameter to normalise the structure function properly, and then
compare our results with the previous ones in the literature.
Correlation effects due to the random phase approximation (RPA) are considered and studied in
Sec.~\ref{sec:RPA}. We present a summary and discussions in
Sec.~\ref{sec:sum}.

\section{Neutrino bremsstrahlung rate: perturbative calculation}
\label{sec:brem}
  
\begin{figure}[htbp]
  \centering
  \plotone{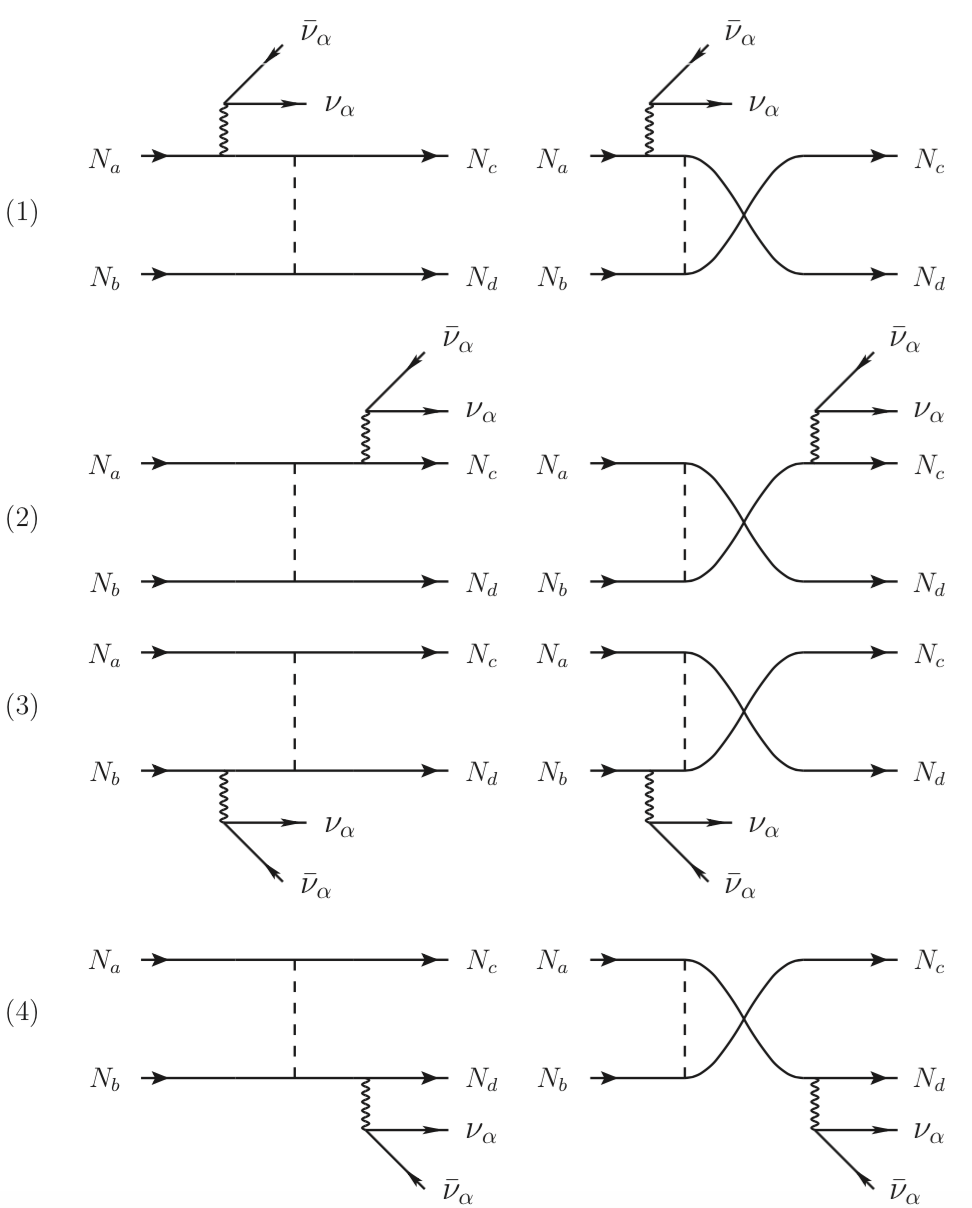}  
  \caption{Feynman diagrams of neutrino bremsstrahlung,
$N_a+N_b\to N_c+N_d+\nu_\alpha+\bar\nu_\alpha$,
where $N$ stands for either a neutron, $n$, or a proton, $p$, and $\alpha=e,\mu,\tau$
for different neutrino flavors. Both the direct (left) and exchange (right)
diagrams are considered. \label{fig:NNB}} 
\end{figure} 

\subsection{Formalism} 
To study neutrino bremsstrahlung and related processes, we consider
the diagrams as shown in Figure~\ref{fig:NNB}, including both
the direct and the exchange contributions. Neglecting weak magnetism
and pseudoscalar corrections, the amplitudes of the diagrams are
\begin{subequations}
  \begin{equation}
  \begin{aligned}
    M^{(1)} &= \frac{G_F C^a_A}{\sqrt{2}}\, {}_{\text{nas}}\langle
    cd|V\bm{\sigma}^{(a)}_i|ab \rangle_{\text{nas}} \,\frac{l_i}{\omega}, \label{eq:Mba}  
    \end{aligned}
  \end{equation}
  \begin{equation}
  \begin{aligned}
    M^{(2)} &= -\frac{G_F C^c_A}{\sqrt{2}}\, {}_\text{nas}\langle
    cd|\bm{\sigma}^{(c)}_i V|ab \rangle_{\text{nas}} \,\frac{l_i}{\omega}, \label{eq:Mbb}  
   \end{aligned} 
  \end{equation}
  \begin{equation}
  \begin{aligned}
    M^{(3)} &= \frac{G_F C^b_A}{\sqrt{2}}\, {}_\text{nas}\langle
    cd|V\bm{\sigma}^{(b)}_i|ab
    \rangle_{\text{nas}}\,\frac{l_i}{\omega}, \label{eq:Mbc}   
  \end{aligned}
  \end{equation}
  \begin{equation}
  \begin{aligned}
    M^{(4)}&=-\frac{G_F C^d_A}{\sqrt{2}}\, {}_\text{nas}\langle
    cd|\bm{\sigma}^{(d)}_i V |ab \rangle_{\text{nas}} \,\frac{l_i}{\omega}, \label{eq:Mbd}  
  \end{aligned}  
  \end{equation}
\end{subequations}
where $|ab\rangle_{\text{nas}}$ and $|cd\rangle_{\text{nas}}$ are 
normalized antisymmetric states of the initial and final nucleon pair,
which are characterized by their relative momenta and spin projections
(see Appendix \ref{sec:app_TsqB} for more details).
For neutrino pair absorption we have
$\omega=(E_{\nu}+E_{\bar\nu})>0$. $\bm{\sigma}_i^{(a)}$ ($i=x,y,z$)
are the Pauli matrices acting on the nucleon $N_a$, $l_i$ are the
spatial components of the leptonic weak current,
$C^a_A=g_A/2\simeq 0.63$ if $N_a$ is a proton and $C_A^a =-g_A/2$ if
it is a neutron, and $G_F$ is the Fermi coupling constant. Note that
we do not include the vector terms in ${\cal M}^{(1,2,3,4)}$ since
they cancel each other out in the non-relativistic
limit within the Born
approximation~\citep{Friman.Maxwell:1979,Raffelt.Seckel:1995,Hannestad.Raffelt:1998}. Going
  beyond the Born approximation and using the half-off-shell
  $T$-matrix, the complete cancellation of the vector terms does not
  hold any longer. Nevertheless, its contribution is negligible compared
  to the axial-vector terms. $V$ could denote either the
nucleon-nucleon scattering $T$-matrix based on the $\chi$EFT potential
of \citet{Entem.Machleidt.Nosyk:2017} with cutoff $\Lambda = 500$~MeV
or the OPE potential. For comparison with previous
results~\citep{Friman.Maxwell:1979,Hannestad.Raffelt:1998,Bartl.Pethick.Schwenk:2014},
the OPE potential is treated in the Born approximation. In the limit
of non-degenerate nucleons our OPE results are identical to those 
of~\citet{Bartl.Pethick.Schwenk:2014}. We will use $\mathcal{T}$ to
explicitly denote the scattering $T$-matrix based on the $\chi$EFT
potential.  \citet{Bartl.Pethick.Schwenk:2014} have shown, that in the
Born approximation, OPE and $\chi$EFT potentials give similar rates at
subsaturation densities as they are dominated by the long-range part of
the tensor force, that is well described by the OPE
potential. However, they also showed that the low-energy resonant
nature of the nucleon-nucleon
interaction~\citep{Bartl.Pethick.Schwenk:2014} enhances the rates and
requires one to go beyond the Born approximation.

The total amplitude can be written in a more compact form as
\begin{equation}
\begin{aligned}
\mathcal{M}_\text{tot}=&\sum_{j=1}^4 M^{(j)} = -\frac{G_Fg_A}{2\sqrt{2}\omega} \\
& \times 
{}_{\text{nas}}\langle cd|[V, \sum_r\bm{\sigma}^{(r)}_i \bm{\tau}_z^{(r)}]'|ab\rangle_{\text{nas}}\, l_i,   
\end{aligned}
\end{equation}  
where $\bm{\tau}_z$ is the $z$-component of the isospin operator with $\bm{\tau}_z|n\rangle=|n\rangle$ and 
$\bm{\tau}_z|p\rangle=-|p\rangle$, and $r$ runs over the two initial or
final nucleons. The prime in the commutator denotes 
  that the potential is evaluated at different values of the energy
  for the first (``positive'') and second (``negative'') terms; see the
  definition in Equation~(\ref{eq:comm}). For energy-independent
  potentials,  such as the OPE or the chiral potential at the Born level, it
  reduces to the standard commutator. The squared amplitude can be
divided into leptonic, $L$, and hadronic, $H$, parts as
$\sum_\text{spins} |\mathcal{M}_\text{tot}|^2 =
(G_F^2g^2_A/8)\sum_{ij} H_{ij}L^{ij}$. For an isotropic medium, we
only need to consider the trace average of the hadronic
part~\citep{Hannestad.Raffelt:1998}:

\begin{equation}
\begin{aligned}
  \label{eq:Htrace}
  \sum_\text{spins} & |\mathcal{M}_\text{tot}|^2 \xrightarrow{\text{isotropic}} G_F^2
    g^2_A \bar{H} E_\nu E_{\nu'} (3 -\cos \theta_{\nu\nu'}), \\
\bar{H} &\equiv \frac{1}{3} \sum_i H_{ii} \\
 & = \frac{1}{3} \sum_{i} \sum_\text{spins}
|{}_\text{nas}\langle cd|[V, \frac{1}{\omega}\sum_{r=1,2}\bm{\sigma}^{(r)}_i \bm{\tau}_z^{(r)}]'|ab\rangle_{\text{nas}}|^2. 
\end{aligned}
\end{equation}
Note that $\bar{H}$ is a scalar under rotations due to the invariance
of the trace under basis transformations. The partial wave expansions
of $\bar{H}^{(nn)}=\bar{H}^{(pp)}$ and $\bar{H}^{(np)}=\bar{H}^{(pn)}$
are presented in Appendix~\ref{sec:app_TsqB}. The calculation of
$\bar{H}$ requires the evaluation of the $T$-matrix elements in
momentum space $\langle \bm{k}_f|\mathcal{T}|\bm{k}_i\rangle$, where
$\bm{k}_i\equiv(\bm{k}_a-\bm{k}_b)/2$ and
$\bm{k}_f\equiv(\bm{k}_c-\bm{k}_d)/2$ are the relative momenta of the
initial and final nucleon pair, and only the initial or final nucleon
pair is on-shell for finite values of $\omega$ (see
Figure.~\ref{fig:NNB}), i.e., we deal with half-off-shell matrix elements.
Here, we fully 
consider their contribution when we solve the Lippmann-Schwinger
equation based on the $\chi$EFT potential
of~\citet{Entem.Machleidt.Nosyk:2017}. Additionally, we include
in-medium Pauli blocking effects (see Appendix~\ref{sec:app_tmatrix})
when solving the Bethe-Goldstone equation. With such medium effects
taken into account, $\bar H$ is a function of
$K=|\bm{K}|\equiv |\bm{k}_a+\bm{k}_b| = |\bm{k}_c+\bm{k}_d|$,
$k_i=|\bm{k}_i|$, $k_f=|\bm{k}_f|$, and $\cos\theta$, where $\theta$ is
the angle between $\bm{k}_i$ and $\bm{k}_f$. 

The response of a nuclear medium can be described by the so-called
structure function or response function. For neutrino bremsstrahlung in
the long-wavelength limit (i.e., we ignore momentum
exchange\footnote{Based on the OPE potential, we have estimated that
the long-wavelength limit introduces an error of $\lesssim 10$\%
at the saturation density.}), the axial structure function, $S^{(\lambda\eta)}_\sigma$,
with $\lambda,\eta=n$ or $p$, is given by \citep{Hannestad.Raffelt:1998,Lykasov.Pethick.Schwenk:2008,Bartl.Pethick.Schwenk:2014}    
\begin{equation}   
\label{eq:S}  
\begin{aligned}
S_\sigma^{(\lambda\eta)}(\omega) =& {1\over n_B} \int
\Biggl(\prod_{l=a,b,c,d} \frac{d^3k_l}{(2\pi)^3}\Biggr) f_af_b(1-f_c)(1-f_d) \\ 
\times & \delta^{(3)}(\bm{k}_a + \bm{k}_b - \bm{k}_c - \bm{k}_d)
\bar H^{(\lambda\eta)}(K, k_i, k_f, \cos\theta) \\
\times & (2\pi)^4 \delta(E_a+E_b-E_c-E_d + \omega),  
\end{aligned}     
\end{equation}
where $n_B$ is the total baryon number density and $f_l$ 
are the Fermi functions. 
Throughout this work,
we always take the non-relativistic energy-momentum relation,
and correspondingly the non-relativistic chemical potential without 
including the rest mass. Note that, unlike the
formalisms adopted in \cite{Hannestad.Raffelt:1998}, we do not need to
consider a symmetry factor for identical nucleon species since our matrix element
is calculated for normalised antisymmetric nucleon states. In the
perturbative limit of Equation~\eqref{eq:S} the total axial structure function, $S_\sigma$, is simply 
\begin{equation}
S_\sigma(\omega)=S^{(nn)}_\sigma(\omega)+S^{(pp)}_\sigma(\omega)+
S^{(np)}_\sigma(\omega). 
\label{eq:Stot}
\end{equation}

The structure function in Equation~(\ref{eq:S}) with the Fermi distributions
and blocking involves a multidimensional integral, which can only be computed numerically. We choose $\bm{k}_i$
along the $z$-axis, and without loss of generality we set $\bm{k}_a$ in
the $xz$-plane with a polar angle denoted by $\theta_a$. We further
denote the polar and the azimuthal angles of $\bm{k}_f$ by $\theta$ and
$\phi$. Once $k_i$, $k_a$, $\theta$, $\phi$, and
$\theta_a$ are specified, all momenta are then fixed, making
Equation~(\ref{eq:S}) a five-dimensional integral. We use the Vegas
subroutine in the CUBA library \citep{Hahn:2005}, invoking a Monte Carlo
algorithm to evaluate all the multidimensional integrals in this work.

In the non-degenerate limit, we have $f_af_b(1-f_c)(1-f_d)
\simeq f_af_b\simeq \exp\{[(-K^2/4-k_i^2)/m_N+\mu_a+\mu_b]/T\}$, independent
of all the angles, and $S_\sigma(\omega)$ can be simplified to
\begin{equation}
\begin{aligned}
    S^{(\lambda\eta)}_{\sigma,m,b}(\omega)=& {4m_N \over
          (2\pi)^5 n_B} \int dKdk_i K^2 k_i^2 k_f  
           \bar H^{(\lambda\eta)}_{L=0}(K, k_i, k_f)\\
            & \times \exp\left\{-\frac{K^2/(4m_N) + k_i^2/m_N - \mu_a-\mu_b}{T}\right\}, 
\end{aligned} 
\end{equation}
where $m_N=(m_n+m_p)/2$ is the averaged nucleon mass, and $\mu_{a,b}$
are the non-relativistic chemical potentials of nucleons. Since
$\int d\cos\theta P_L(\cos\theta)=2\delta_{L0}$, where $P_L$ is the
Legendre polynomial, only the $L=0$ component of $\bar H$ contributes;
see Equation~(\ref{eq:Hsum}).  We use the subscript $m$ ($v$) to refer to
the in-medium (vacuum) $T$-matrix elements, and $b$ ($f$) when we use the
Boltzmann (Fermi) distribution without (with) blocking.\footnote{Not to
  be confused with the medium blocking for the $T$-matrix.  From now
  on, we will always use ``blocking'' to refer to the Pauli blocking of the
  final nucleon states as shown in Equation~(\ref{eq:S}), unless otherwise
  specified.} Throughout this work, we always take the bare nucleon
mass for all our studies. For typical densities in the
neutrinosphere, the effective mass of nucleons is close to the bare
value. At the saturation density, $\chi$EFT
calculations~\citep{Hebeler.Duguet.ea:2009,Wellenhofer.Holt.ea:2014,Drischler.Krueger.ea:2017}
found an effective mass $\sim 0.9\,m_N$. Using such a value for both
proton and neutron, the rates are only affected by a few per cent.

When the vacuum $T$-matrix elements or the OPE potential are used,
$\bar{H}^{(\lambda\eta)}$ is independent of $K$ and integration over
$K$ can be done analytically with
$\int dKK^2 \exp(-K^2/(4m_NT)) = \sqrt{4\pi (m_NT)^3}$, leading to
\begin{equation}
\begin{aligned}
\label{eq:S_nondeg}        
S^{(\lambda\eta)}_{\sigma,v,b}(\omega)  
 =& {(2m_N)^{5/2} T^{3/2} \over (2\pi)^{9/2} n_B} \int
    dk_i k_i^2 k_f \bar H^{(\lambda\eta)}_{L=0}(k_i, k_f) \\  
& \times \exp\left\{-\frac{k_i^2/m_N - \mu_a-\mu_b}{T}\right\}. 
\end{aligned}    
\end{equation}  

Once $S_\sigma(\omega)$ is known, the inverse mean free path or opacity of
a neutrino against neutrino pair absorption is

  \begin{equation}
    \label{eq:lambda_a}     
    \begin{aligned}
      \lambda^{-1}_{A}(E_{\nu}) =& \frac{g_A^2 G_F^2 n_B}{4} \int {d^3k' \over
                                   (2\pi)^3} (3-\cos\theta_{\nu\bar\nu}) f'(E'_\nu) \\ 
                                   & \times S_\sigma(E_\nu+E'_\nu),
    \end{aligned}
  \end{equation}
where $\bm{k}'$ and $f'$ are the momentum and distribution function of the
counterpart (anti)neutrino, and $\theta_{\nu\bar\nu}$ is the angle between the neutrino momenta. 

The spectrum of emitted neutrinos with a particular flavor per unit
of solid angle is
\begin{equation}
\begin{aligned}
\label{eq:nu_emi}   
\phi(E_\nu) = & \frac{g_A^2 G_F^2 n_B E_\nu^2}{4 (2\pi)^3} \int {d^3k' \over (2\pi)^3}
(3-\cos\theta_{\nu\bar\nu})\\
& \times S_\sigma(-E_\nu-E_{\nu'}) [1-f(E_\nu)][1-f'(E_{\nu'})].  
\end{aligned}  
\end{equation}
If one can neglect the final-state blocking of neutrinos, Equation~(\ref{eq:nu_emi}) can be further simplified to

\begin{equation}
\begin{aligned}
 \label{eq:nu_emi_nob} 
 \phi(E_\nu) \approx  \frac{3 g_A^2 G_F^2 n_B E_\nu^2}{4(2\pi)^3} & \int {4\pi E_{\nu'}^2 dE_{\nu'}
   \over (2\pi)^3} S_\sigma(E_\nu+E_{\nu'})  \\ 
  & \times e^{-(E_\nu+E_{\nu'})/T},  
\end{aligned}
\end{equation} 
where we have used the detailed balanced relation 
$S_\sigma(-\omega)=S_\sigma(\omega)\exp(-\omega/T)$. Assuming thermal
distributions for neutrinos, we find
$\phi(E_\nu)/\lambda_A^{-1}(E_\nu) \propto E_\nu^2 \exp(-E_\nu/T)$. For demonstration, we always use Equation~(\ref{eq:nu_emi_nob}) to calculate the neutrino spectra emitted, but the final-state neutrino blocking can be easily included in neutrino transport in realistic supernova simulations.
Similarly, the energy loss rate due to neutrino pair emission is
\begin{equation}
\label{eq:Qbrem} 
\begin{aligned}
Q_{\text{brem}} =& \frac{3}{4}g_A^2 G_F^2 n_B \int {d^3k \over (2\pi)^3}  {d^3k'
  \over (2\pi)^3} (3-\cos\theta) (E_\nu+E'_\nu), 
\end{aligned}
\end{equation}
which can be approximated by 
\begin{equation}
\label{eq:Qbrem_nob} 
\begin{aligned}
 Q_{\text{brem}}
 \approx &\frac{3 g_A^2 G_F^2 n_B}{160\pi^4} \int_0^\infty d\omega
           \omega^6 e^{-\omega/T} S_\sigma(\omega), 
\end{aligned}
\end{equation}
if the final-state neutrino blocking can be neglected. Note that the prefactor 3 accounts for three different neutrino flavors.

Assuming the neutrino spectrum follows a Boltzmann distribution,
$f(E_\nu) \propto \exp(-E_\nu/T_\nu)$, with a neutrino temperature
$T_\nu=T$, the energy-averaged pair absorption inverse mean free path
per neutrino can be expressed as
\begin{equation}
  \label{eq:mfp_a}
  \begin{aligned} 
    \frac{\langle \lambda^{-1}_A \rangle}{n_\nu'} 
    \equiv& \frac{1}{n'_\nu} \frac{\int \lambda_A^{-1}(E_\nu) f(E_\nu)
      E_\nu^2 dE_\nu}{\int f(E_\nu) E_\nu^2 dE_\nu}  \\      
    =&  \frac{g_A^2 G_F^2 n_B}{160 T^6}
    \int_0^\infty d\omega \omega^5
    e^{-\omega/T}
    S_\sigma(\omega).
  \end{aligned}                                    
\end{equation}
We use the inverse mean free path per
neutrino number density instead of the inverse mean free path because the latter depends
on the number density of neutrinos, which needs to be determined by full
Boltzmann transport calculations.

\subsection{Energy-averaged inverse mean free path using different treatments} 
\label{sec:unnorm}

As already mentioned above, we can perform calculations based on
different schemes: vacuum $T$-matrix, in-medium $T$-matrix and OPE
potential; each considers different approximations: either
on-shell or half-off-shell and the Boltzmann or Fermi distribution
that includes final-state blocking. In what follows, we firstly
consider $\langle \lambda_A^{-1} \rangle/n_\nu'$ based on the
vacuum/in-medium $T$-matrix and the OPE potential, and explore the
effects of the different approximations. As
in~\citet{Bartl.Pethick.Schwenk:2014}, we take the typical conditions
in SNe characterized by
\begin{align}
T_{SN}(\rho) = 3~\text{MeV}~\Big({\rho \over 10^{11}~\text{g cm}^{-3}}\Big)^{1/3}, 
\label{eq:SN_T}
\end{align} 
and choose $Y_e=0.1, 0.3$, and 0.5 for the following studies.

\begin{figure*}[htb]      
\plotone{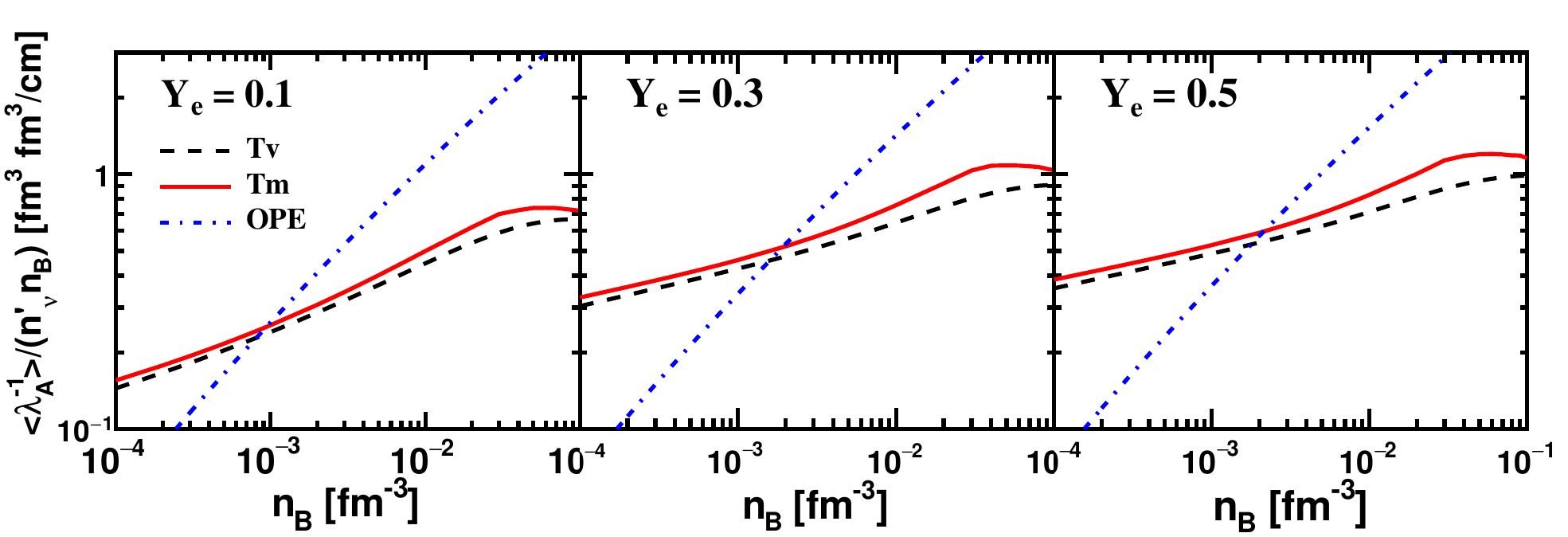}
\caption{$\langle \lambda_A^{-1} \rangle/(n_\nu'n_B)$ as functions of density
with the temperature given by Equation~(\ref{eq:SN_T}) and
$Y_e=0.1, 0.3$ and 0.5, respectively. The half-off-shell elements based on the vacuum $T$-matrix, 
in-medium $T$-matrix, and OPE potential have been used for comparison. \label{fig:opacity_un}}       
\end{figure*}

\begin{figure*}[htb]
\plotone{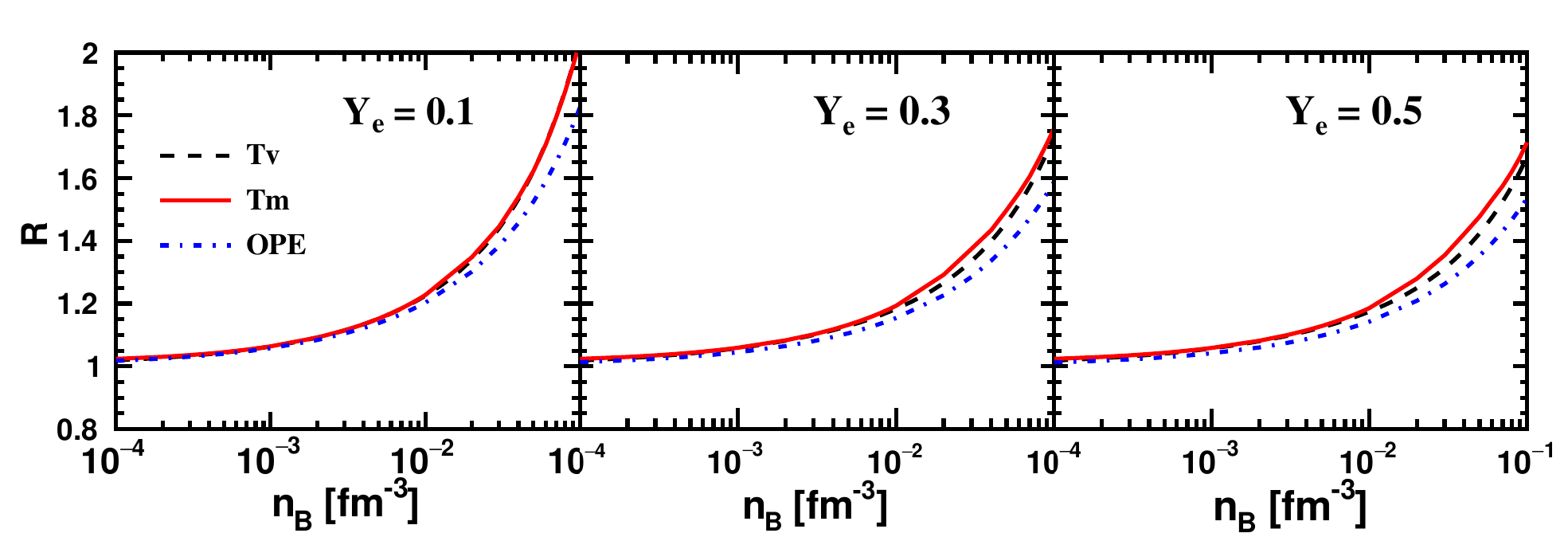}  
\caption{ Ratios of $\langle \lambda_A^{-1} \rangle$ using the 
Boltzmann distribution without blocking to those using the Fermi
distribution with blocking. The half-off-shell elements are used. \label{fig:bvsf}}       
\end{figure*}

Figure~\ref{fig:opacity_un} compares the results of
$\langle \lambda_A^{-1} \rangle/(n_\nu'n_B)$ using the vacuum
$T$-matrix, in-medium $T$-matrix, and the OPE potential. By dividing by
the explicit factor $n_B$ in Equation~(\ref{eq:mfp_a}), the value of
$\langle \lambda_A^{-1} \rangle/(n_\nu'n_B)$ still increases with
density as shown in Figure~\ref{fig:opacity_un} due to the temperature 
dependence of Equation~(\ref{eq:SN_T}), which results in neutrinos with higher
energies as the density grows. This will be further discussed when the
normalized structure function is introduced.

Compared to the OPE potential, the $T$-matrix leads to an enhancement
of $\langle \lambda_A^{-1} \rangle$ below $\sim$ 0.001--0.002
fm$^{-3}$, i.e., $\rho \sim (1.7\text{--}3.4)\times 10^{12}$ g cm$^{-3}$,
and a suppression
above.
The enhancement at low densities for the $T$-matrix is due to the
resonant property of the nuclear force~\citep{Bartl.Pethick.Schwenk:2014}.
At high densities, higher relative momenta become more relevant for
which the $T$-matrix elements are suppressed and hence the inverse
mean free path. Medium effects on the $T$-matrix lead to a slight increase in the
bremsstrahlung rate by $\sim 10\%$. The effect is relatively small because for
the conditions we consider in Equation~(\ref{eq:SN_T}) nucleons are not very degenerate, and 
meanwhile, the effects on the real and the imaginary parts of the $T$-matrix 
balance with each other. We choose to show the in-medium $T$-matrix results for 
the following studies but will not focus on the details. 

\begin{figure*}[htb]               
\plotone{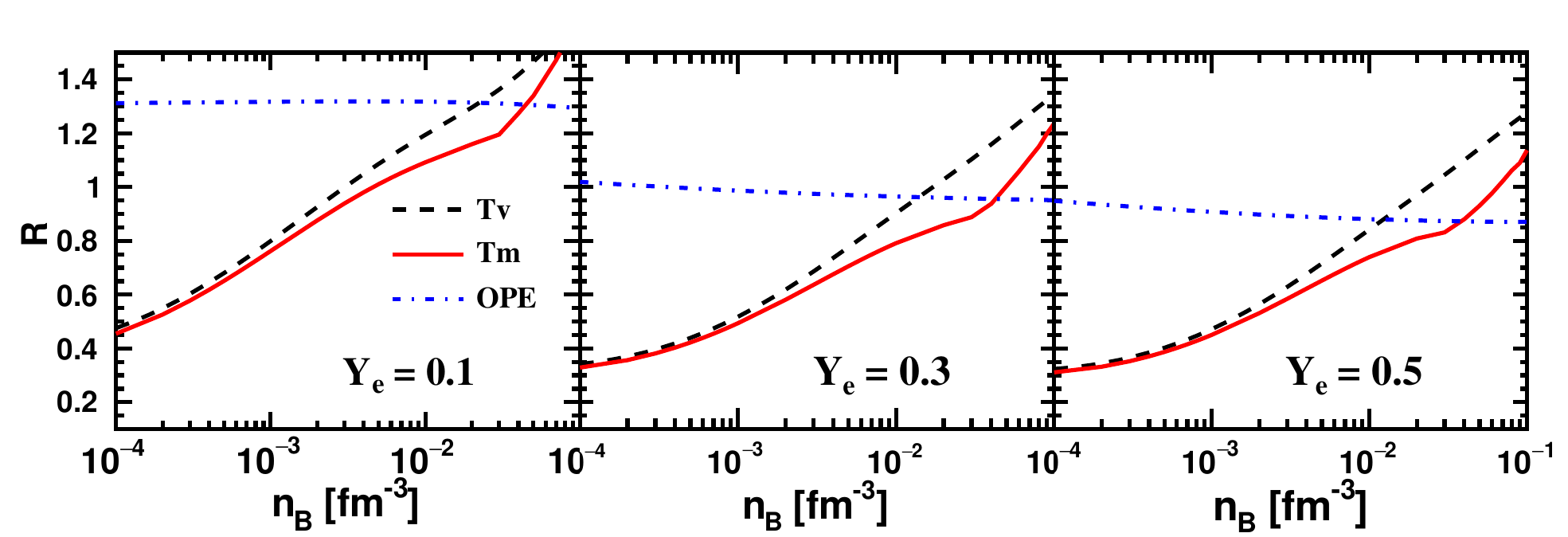} 
\caption{ Ratios of $\langle \lambda_A^{-1} \rangle$ using the 
  effective on-shell diagonal elements to those based on the half-off-shell/non-diagonal
  elements. The Fermi distribution with blocking is used. \label{fig:diag}}     
\end{figure*}

Figure~\ref{fig:bvsf} shows the ratios of
$\langle \lambda_A^{-1} \rangle$ using the Boltzmann distributions without
blocking to those using the Fermi distributions with blocking,
where the half-off-shell elements are used for both cases. 
The impact of blocking increases with density as the nucleon
degeneracy increases. Using Equation~(\ref{eq:SN_T}), the degeneracy
parameter for neutrons can be expressed as
$\epsilon_n = E_F^{(n)}/T \simeq 0.1 [\rho(1-Y_e)/10^{11}~\text{g
  cm}^{-3}]^{1/3} \simeq 0.12 [n_B(1-Y_e)/10^{-4}\
\text{fm}^{-3}]^{1/3}$. We find that the Boltzmann approximations
overestimate the opacity by $\sim$ 20\% at
$\epsilon_n \simeq 0.5$, i.e., at $n_B\simeq 10^{-2}\ \text{fm}^{-3}$,
and by $\sim$ 50\%--100\% at $n_B\simeq 10^{-1}\ \text{fm}^{-3}$. As expected,
the impact of the Pauli blocking is insensitive to 
the nuclear potentials used.

As given in Equations~(\ref{eq:phase_bar}) and (\ref{eq:phase_bar_uncoup}),
the on-shell diagonal vacuum $T$-matrix is related to the
experimentally measured phase shifts and mixing parameters; see Appendix \ref{sec:app_tmatrix}. This provides a method to estimate the
on-shell diagonal elements of the
$T$-matrix. Following~\cite{Bartl.Pethick.Schwenk:2014}, we use an
effective on-shell element
$\langle \bar k | \mathcal{T} | \bar k \rangle$ to approximate the
half-off-shell and non-diagonal $T$-matrix element
$\langle k_f|\mathcal{T}|k_i\rangle$ with
$\bar k = \sqrt{(k_i^2 + k_f^2)/2}$. This approximation has been found to be
reasonable for the OPE potential~\citep{Bartl.Pethick.Schwenk:2014}.
Compared to the studies based on the half-off-shell $T$-matrix, the effective on-shell matrix elements
underestimate the rates significantly for densities
$n_B\lesssim 10^{-2}$~fm$^{-3}$ by a factor up to $0.7$, and
overestimate them above (see Figure~\ref{fig:diag}). Therefore, the
use of the half-off-shell $T$-matrix is required to reach an accurate
bremsstrahlung rate.
     
\section{Normalized structure function}   
\label{sec:norm}  

The structure function, $S_\sigma(\omega)$, given in Equation~(\ref{eq:S}) in
the long-wavelength limit diverges as $\omega^{-2}$ for $\omega\to 0$,
which is a common feature of any bremsstrahlung-type
process~\citep{Raffelt.Seckel.Sigl:1996}.
Though there is no divergence for the inverse mean free path
$\langle \lambda^{-1}_A \rangle$ studied in Sec.~\ref{sec:unnorm}, it
may lead to an unphysical enhancement of $\lambda^{-1}_A(E_\nu)$ in
the limit of $E_\nu \to 0$. Hence, we want to obtain a well-behaved
$S_\sigma(\omega)$ and study how the related rates are modified. It
also provides a proper comparison with the existing studies with
well-behaved structure
functions~\citep{Hannestad.Raffelt:1998,Raffelt:2001,Bartl.Pethick.Schwenk:2014}. This
also allows us to extend the calculations to include RPA correlation
effects based on a smooth $S_\sigma(\omega)$, as will be done in  
Sec.~\ref{sec:RPA}.  

It has been suggested~\citep[see,
e.g.,][]{Hannestad.Raffelt:1998,Raffelt:2001,Lykasov.Pethick.Schwenk:2008,Bacca.Hally.ea:2009,Bacca.Hally.ea:2012,Roberts.Reddy.Shen:2012,Bartl.Pethick.Schwenk:2014,Roberts.Reddy:2017}
that the structure function can be regularized by replacing $\omega^{-2}$
with $(\omega^2+\Gamma^2)^{-1}$, where the width parameter $\Gamma$ is
introduced to characterize the spin fluctuation or relaxation
rate. The axial structure function can also be viewed as a spin 
autocorrelation function, which is expected to decay exponentially
as $\exp(-\Gamma t)$ at long times, leading to a Lorentzian form
of
$S_\sigma(\omega)$~\citep{Hannestad.Raffelt:1998,Raffelt:2001}. This
is equivalent to considering that the nucleon propagator has a width due
to nucleon-nucleon scattering in the nuclear medium, i.e., replacing $\omega^{-1}$
by $(\omega+i\Gamma)^{-1}$. Therefore, the proper renormalization of
nucleons in the medium \citep[also called `multiple-scattering' effects in the
literature, see, e.g.,][]{Hannestad.Raffelt:1998} renders $S_\sigma(\omega)$ a
well-behaved function. Studies based on Landau's Fermi liquid theory that
compute an energy-dependent relaxation rate also lead to a
well-behaved
$S_\sigma(\omega)$~\citep{Lykasov.Pethick.Schwenk:2008,Bacca.Hally.ea:2009,Bacca.Hally.ea:2012,Bartl.Pethick.Schwenk:2014,Bartl:2016}.
Since the relaxation rate varies very slowly with $\omega$, we find
that $S_{\sigma}(\omega)$ regularized by a constant
$\Gamma$ agrees within a few per cent with the results
of~\citet{Bartl.Pethick.Schwenk:2014}.

\begin{figure*}[htbp]              
\plotone{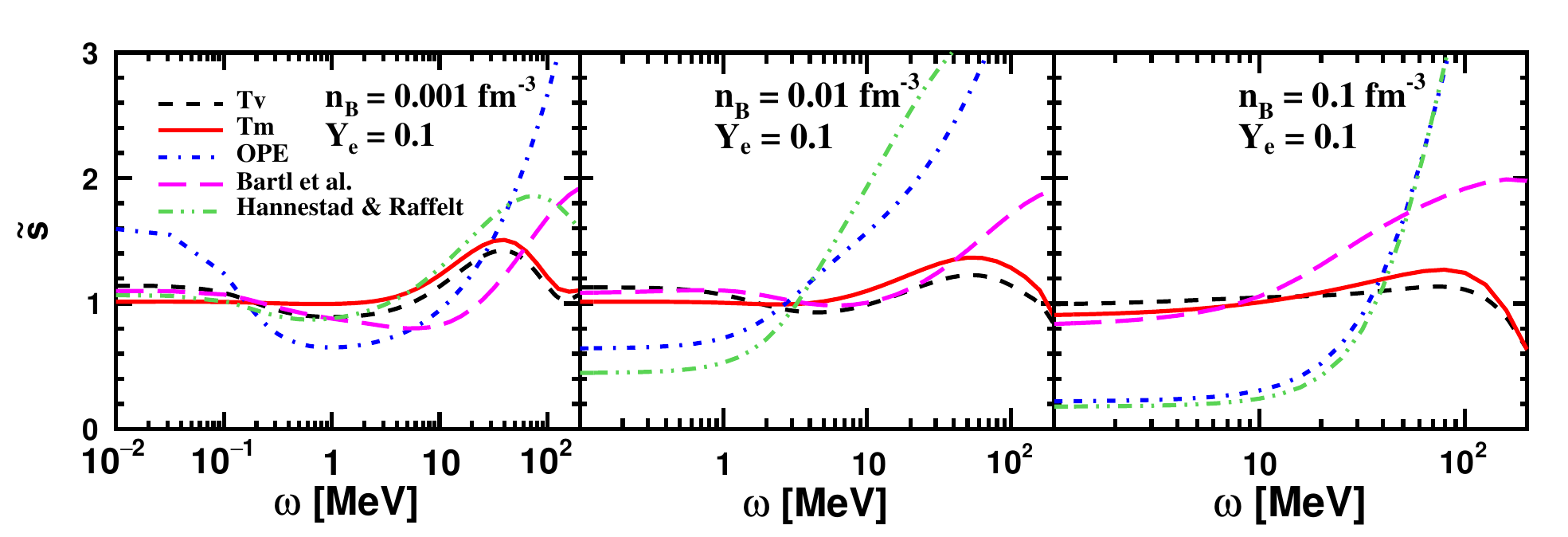}          
\caption{$\tilde s$ as functions of $\omega$ at $n_B=0.001$~fm$^{-3}$, 0.01
fm$^{-3}$ and 0.1 fm$^{-3}$, respectively, for $Y_e = 0.1$. The
half-off-shell matrix elements of the vacuum $T$-matrix, in-medium $T$-matrix,
and OPE potential with the Pauli blocking are used. $\tilde s(\omega)$
from the fitting formulae in \citet{Hannestad.Raffelt:1998} based on the OPE
potential, and obtained using the effective on-shell $T$-matrix following the
formalism of \citet{Bartl.Pethick.Schwenk:2014}, are also added 
for comparison. \label{fig:s_tilde}}  
\end{figure*} 

The parameter $\Gamma$ can be determined by the normalization
condition \citep{Raffelt.Strobel:1997,Hannestad.Raffelt:1998}:  
\begin{equation}
  \label{eq:norm}
\begin{aligned}  
\int_{-\infty}^\infty {d\omega \over 2\pi} S_\sigma(\omega) &= \int_0^\infty {d\omega \over 2\pi} S_\sigma(\omega) (1+e^{-\omega/T}) \\
&={2\over n_B}
  \sum_{i=n,p}\int {d^3k\over (2\pi)^3} f_i(\varepsilon(k)) [ 1-
  f_i(\varepsilon(k))],    
\end{aligned}   
\end{equation} 
with $\varepsilon(k)=k^2/(2m_N)$ and $n_B$ the total nucleon number
density.  Note that the above equation is exact for a non-interacting
system, and we assume that the main effect of nucleon-nucleon
collisions is to increase the width of $S_\sigma(\omega)$ while
keeping the normalization. Unless otherwise stated, $S_\sigma(\omega)$ refers to the properly normalized
structure function, and we call the ones computed in Equations~(\ref{eq:S})-(\ref{eq:S_nondeg}) unnormalized structure functions.

\begin{figure*}[htbp]     
\plotone{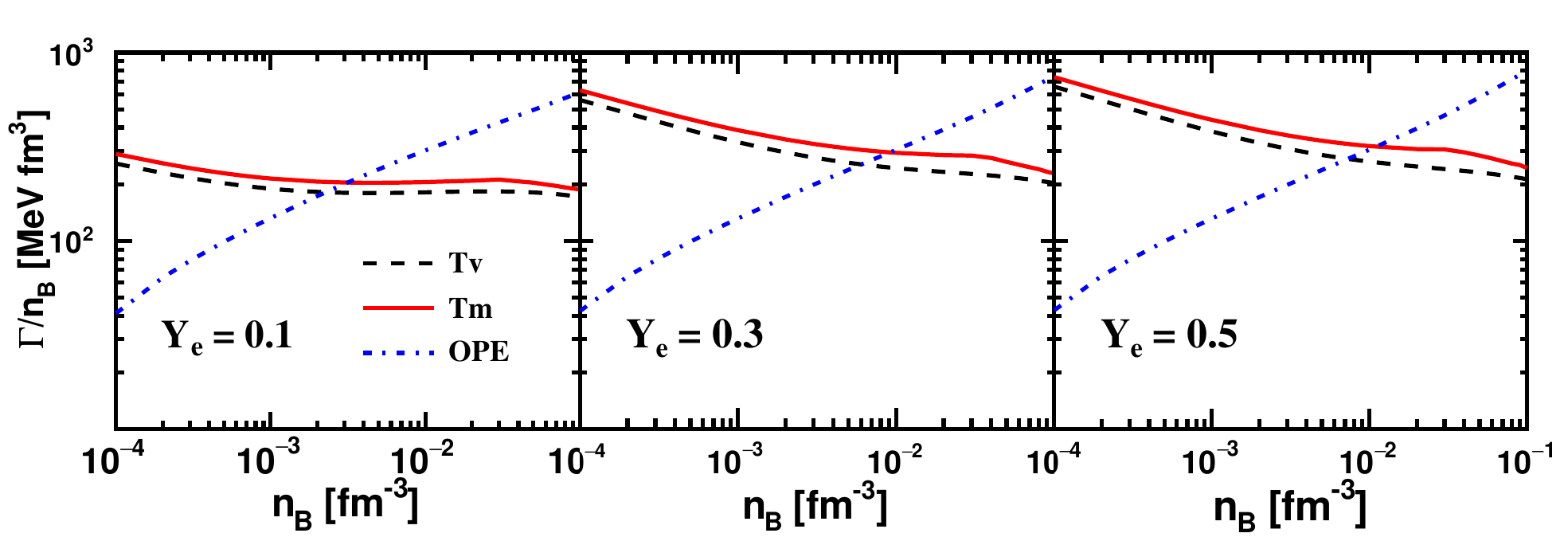} 
\caption{$\Gamma/n_B$ as a function of density based on different nuclear
matrix elements for $Y_e = 0.1$, 0.3, and 0.5,
respectively.\label{fig:Gamma}}
\end{figure*}   
 
%\todo[inline]{GMP: I am here}
  
The normalized $S_\sigma(\omega)$ can be expressed in Lorentzian form as~\citep{Hannestad.Raffelt:1998,Raffelt:2001,Lykasov.Pethick.Schwenk:2008,Bacca.Hally.ea:2009,Bacca.Hally.ea:2012,Roberts.Reddy.Shen:2012,Bartl.Pethick.Schwenk:2014,Roberts.Reddy:2017}
\begin{equation}
  \label{eq:smalls}
  S_\sigma(\omega) = s(\omega)\frac{2\Gamma}{\omega^2+\Gamma^2},
\end{equation}
where $s(\omega)$ is a dimensionless quantity that contains additional
energy dependences originating from the nuclear correlations and
blocking. For $\omega \ll \Gamma$, $s(\omega) \simeq 1$, and one has
$S_\sigma(\omega) \simeq 2/\Gamma$; for $\omega \gg \Gamma$,
$S_\sigma \simeq 2s(\omega)\Gamma/\omega^2$, which is fully determined
by the perturbative calculation in Equations~(\ref{eq:S}) and
(\ref{eq:Stot}). 
 
Taking as a reference the calculation based on the in-medium
$T$-matrix, we introduce
\begin{equation}
\tilde s(\omega)\equiv S_\sigma(\omega)\frac{\omega^2+\Gamma_{T_m}^2}{2\Gamma_{T_m}} 
= \frac{S_\sigma(\omega)}{S^{T_m}_\sigma(\omega)} \tilde s^{T_m}(\omega),  
\end{equation}  
with $\Gamma_{T_m}$, $S_\sigma^{T_m}$
and $\tilde s^{T_m}$ the width and structure functions using the in-medium $T$-matrix.
We present the comparison of $\tilde s(\omega)$ in 
Figure~\ref{fig:s_tilde} to show the relative differences in $S_\sigma(\omega)$
when using different approaches.
Results based on the fitting formulae of the structure function from
\citet{Hannestad.Raffelt:1998} consider only neutron-neutron
interactions using the OPE potential. To
demonstrate the effects of the off-shell elements and blocking, we
also show the results based on the effective on-shell vacuum $T$-matrix
following the formalism of~\citet{Bartl.Pethick.Schwenk:2014}, where
the blocking effects are neglected and
$S_\sigma(\omega)$ is
normalized to $\sim$ 1; see Equation~(\ref{eq:norm}).

At low density condition where the blocking of the final nucleons can
be ignored (see the left panel of Figure~\ref{fig:s_tilde}), we find
an underestimation of $\tilde s(\omega)$, or $S_\sigma(\omega)$, at
intermediate $\omega$ and an overestimation for high
$\omega \gtrsim 20$ MeV, when the effective on-shell $T$-matrix
elements are used. This is also consistent with the results shown in
Figure \ref{fig:diag}, considering that $S_\sigma(\omega)$ for
$\omega \sim 3 T$ dominates the inverse mean free path, see
Equation~(\ref{eq:mfp_a}).  As density increases, the Pauli blocking starts
to play a role, and its impact becomes comparable to or even dominant
over that of off-shell effects (see the middle and the right
panels). For $\omega \gtrsim 10$~MeV, the off-shell effects and
blocking together suppress $S_\sigma$ significantly. It is also
interesting to notice that $s(\omega)$ based on the half-off-shell
$T$-matrix including the blocking is close to 1 with a maximum
deviation of $\sim$ 40\%.  For $s(\omega) \sim 1$, $S_\sigma(\omega)$
based on the $T$-matrix is simply determined by the width parameter; see
Equation~(\ref{eq:smalls}).

The behavior of $S_\sigma(\omega)$ for high values of $\omega$
is very important to the energy-averaged opacity against pair
absorption (and bremsstrahlung energy loss rate) due to the factor of
$\omega^5$ ($\omega^6$) in the integral [see Equations~(\ref{eq:mfp_a}) and (\ref{eq:Qbrem})]. As will be shown later, the combined effect of
off-shell elements and blocking on $S_\sigma(\omega)$ at high $\omega$
can give rise to notable differences in
$\langle \lambda_A^{-1}\rangle$.

Compared to the $T$-matrix, the OPE potential gives rise to a very
different $S_\sigma(\omega)$, or $\tilde s(\omega)$. Since
$S_\sigma(\omega \to 0) \simeq 2/\Gamma$, the value of
$\tilde s(\omega)$ at small $\omega$ is determined by the width
parameter $\Gamma$. The resonant property of the
nuclear force exhibited in the $T$-matrix at low density (temperature) will 
lead to a larger $\Gamma$ and thus a smaller $S_\sigma(\omega)$ or
$\tilde s(\omega)$ at small $\omega$. The $T$-matrix elements decrease rapidly with the relative momenta. For high values of $\omega$, the relative momenta between nucleons become large, leading to smaller values of $S_\sigma(\omega)$ than in the OPE results.

Studies by \citet{Hannestad.Raffelt:1998} consider only neutron-neutron
interactions and hence our OPE results
are close to theirs at low $Y_e$. Aside from the relatively small
errors introduced in the fitting formulae of $S_\sigma(\omega)$
in \citet{Hannestad.Raffelt:1998}, the remaining differences are
due to the use of different $\pi NN$ coupling constants. We use
$[g_A/(2F_\pi)]^4$ in calculating the matrix element with $g_A= 1.26$
and $F_\pi= 92.4$ MeV, which is about $\sim 30$\% smaller than $(f/m_\pi)^4$
used in~\citet{Hannestad.Raffelt:1998} with $f= 1$ and $m_\pi$ the pion mass.

The corresponding values of $\Gamma/n_B$ required to normalize
$S_\sigma(\omega)$ are shown in Figure~\ref{fig:Gamma} as a function
of $n_B$. For $n_B$ around 0.01~fm$^{-3}$, $\Gamma$ can be as
high as a few MeV. As already mentioned above, the resonant property
at low energy/density and a rapidly decreasing $T$-matrix element with
relative momenta are responsible for the enhancement/suppression of
$\Gamma$ at low/high density compared to the OPE results. Furthermore, the behaviors of $\Gamma/n_B$ based on the $T$-matrix and the OPE
potential can be understood in a more quantitative way as follows. At low energy, the $T$-matrix is dominated by the two resonant channels, $^1S_0$ and
$^3S_1$, hence the corresponding hadronic part of
the matrix element for $\omega \to 0$, $\bar H(k, k)$, varies with the relative momenta
as
$\bar H(k, k) \propto |T_{S=0,1}|^2 \propto
(a_{S=0,1}^{-2}+k^2)^{-1}$, where $a_{S=0,1}$ are the scattering
lengths. For comparison, the OPE potential leads to a different
hadronic part, which takes a form like
$\bar H(k, k) \propto k^4/( k^2+m_\pi^2)^2$
\citep{Hannestad.Raffelt:1998}. We consider the non-degenerate
conditions where $k \sim \sqrt{T m_N}$, and from the power counting of
$T$ in Equation~(\ref{eq:S_nondeg}) for the unnormalized structure function we find
$\Gamma/n_B \propto \lim_{\omega\to 0}[\omega^2
S^{(0)}_\sigma(\omega)]/n_B \propto T^{1/2} (m_NT + \eta a^{-2})^{-1}$
for using the $T$-matrix, and
$\Gamma/n_B \propto T(m_N T + \eta' m_\pi^2)^{-1/2}$ for using the OPE
potential, with the coefficients $\eta,\eta' \sim {\cal O}(1)$.  Using
the physical values for $a_{S=0,1}$, $m_\pi$ and $m_N$, one can
explain the behavior of $\Gamma/n_B$ with density (or temperature)
based on the $T$-matrix or the OPE potential shown in Figure
\ref{fig:Gamma}.

\begin{figure*}[htbp]   
%\epsscale{0.87}   
\plotone{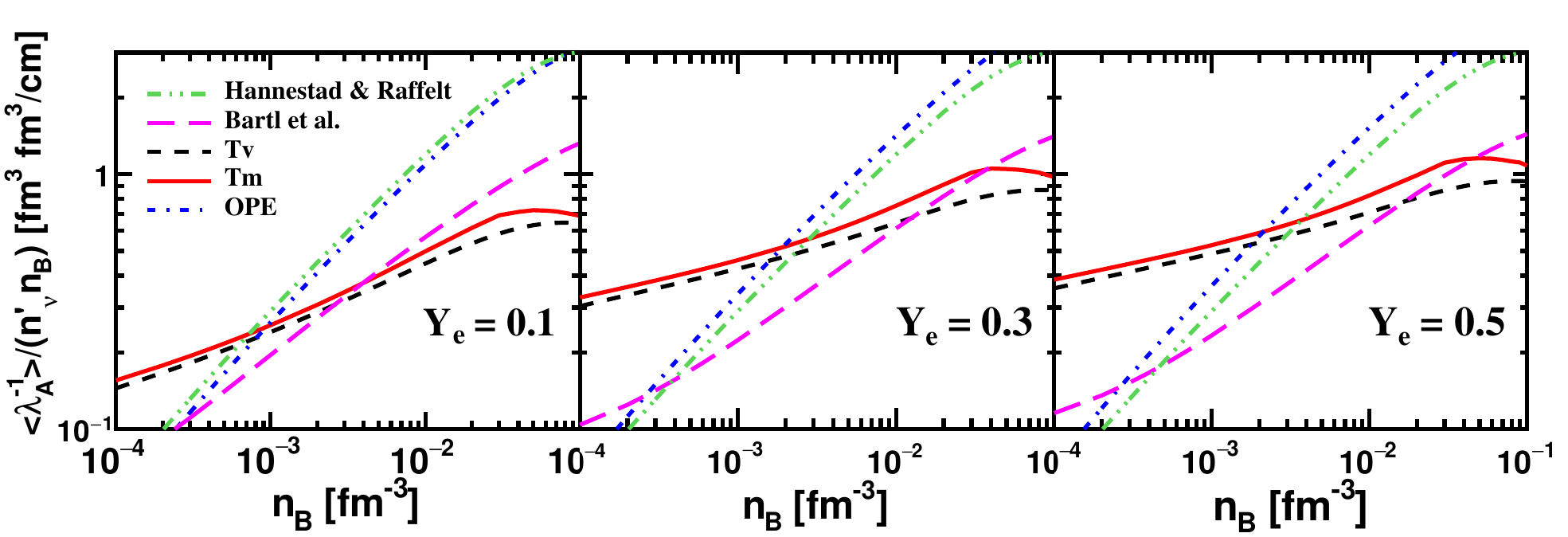}           
\caption{$\langle \lambda_A^{-1} \rangle/(n'_\nu n_B)$ based on normalised $S_\sigma$ as a
function of density for $Y_e = 0.1$, 0.3, and 0.5, respectively.  \label{fig:Opa_absorption}}
\end{figure*}

\begin{figure*}[htbp]  
%\epsscale{0.87}     
\plotone{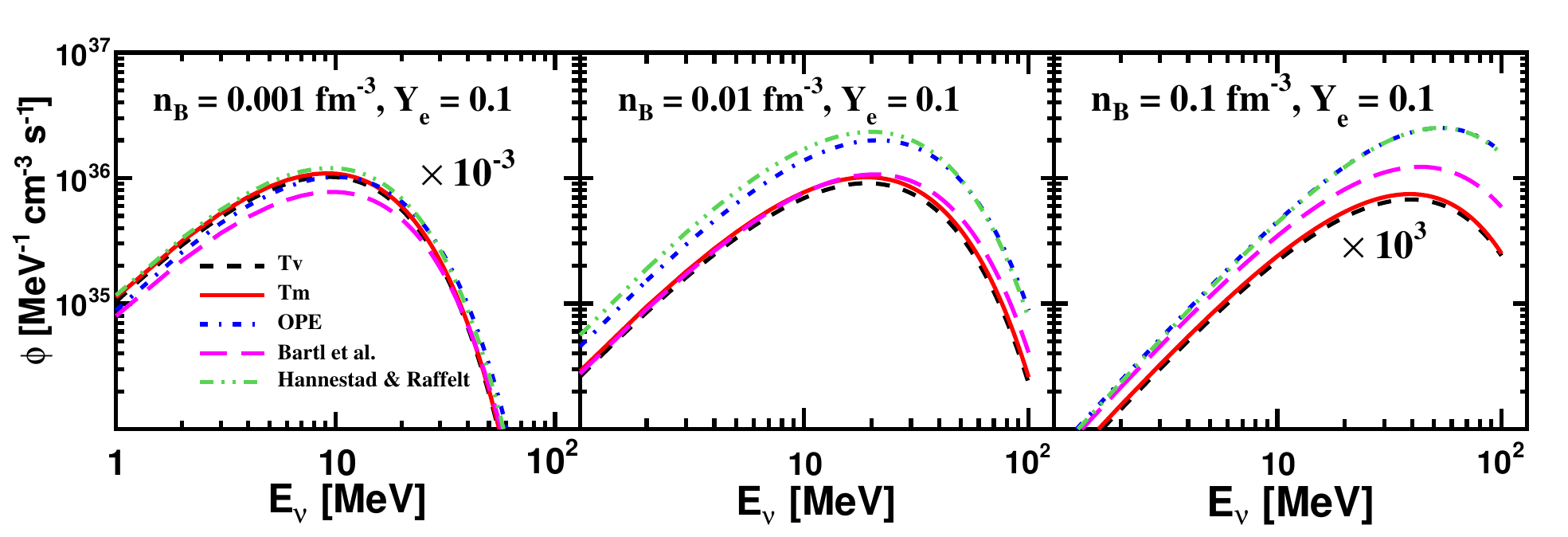}         
\caption{Emissivities of neutrinos from bremsstrahlung $\phi(E_\nu)$, given in Equation~(\ref{eq:nu_emi_nob}), at $n_B = 0.001, 0.01$, and 0.1 fm$^{-3}$, respectively, for $Y_e=0.1$. Note that the results for $n_B = 0.001$ and 0.1 fm$^{-3}$ shown in the plots need to be multiplied by an additional factors of $10^{-3}$ and $10^3$, respectively. \label{fig:emis_E} }                   
\end{figure*}

Figure~\ref{fig:Opa_absorption} compares the results for
$\langle \lambda_A^{-1}\rangle/(n_Bn_\nu')$ based on different
normalized structure functions. The differences are simply due to
different $S_\sigma(\omega)$ at high $\omega$, as shown in
Figure~\ref{fig:s_tilde}.
It should also be emphasized that 
$\langle \lambda_A^{-1}\rangle$ based on the normalized $S_\sigma$ 
are only slightly smaller (by up
to $\sim$ 10\%) than those based on the unnormalized ones; see
Figure~\ref{fig:opacity_un}. Therefore, the studies
of $\langle \lambda_A^{-1}\rangle$ based on the unnormalized structure functions in subsection~\ref{sec:unnorm} still hold.

A more relevant quantity to neutrino transport in supernova matter is
the energy-dependent opacity against pair absorption, $\lambda_A^{-1}(E_\nu)$,
defined in Equation~(\ref{eq:lambda_a}), and the neutrino emissivity
from nucleon-nucleon bremsstrahlung, $\phi(E_\nu)$, given in Equation~(\ref{eq:nu_emi_nob}) neglecting the final-state blocking of neutrinos.
Since $\phi(E_\nu)$ does not depend on the (anti)neutrino number density
if the Pauli blocking is neglected, we choose to show $\phi(E_\nu)$ in 
Figure~\ref{fig:emis_E} at $n_B=$ 0.001, 0.01, and 0.1 fm$^{-3}$,
for $Y_e=0.1$. Note that 
$\lambda^{-1}_A(E_\nu)$ can be obtained simply from $\phi(E_\nu)$ with
$\phi(E_\nu)/\lambda_A^{-1}(E_\nu) \propto E_\nu^2 \exp(-E_\nu/T)$.
At low density (temperature) or for low $E_\nu$, the $T$-matrix with half-off-shell matrix elements 
gives rise to the largest emissivities. As density (temperature) or $E_\nu$
increases, using the effective on-shell $T$-matrix or the OPE potential
overestimates the emissivities. For the range of density and $E_\nu$ explored
in Figure~\ref{fig:emis_E}, the ratios of emissivities based on 
the effective on-shell $T$-matrix and the OPE potential to those based
on the in-medium $T$-matrix with half-off-shell elements range from
$\sim$ 0.5--1.8 and $\sim$ 0.7--5, respectively. 
Just as for the energy-averaged inverse mean free path
$\langle \lambda_A^{-1} \rangle$, the medium effect on the $T$-matrix
increases $\phi(E_\nu)$ by $\lesssim$ (10--20)\%.

We provide a numerical table of the normalized structure function $S_\sigma(\omega)$ based on the vacuum $T$-matrix for calculating the bremsstrahlung rate; see Appendix~\ref{sec:table}. To implement the table in SN simulations, one has to do a 4D interpolation of the structure function over temperature, density, $Y_e$, and $\omega$. It should also be pointed out that, since we use exactly the same notation as that adopted in \cite{Hannestad.Raffelt:1998}, the implementation of our new structure function should be similar.

\section{RPA correlations} \label{sec:RPA} 

In addition to the multiple-scattering effects discussed
above, there is another correlation effect that has been investigated
within the framework of the RPA
\citep{Burrows.Sawyer:1998,Reddy.Prakash.ea:1999}. Multiple-scattering effects account for the
renormalization of the virtual nucleon propagator in the medium, while
  the RPA correlations screen the coupling between the leptonic weak
  current and the nucleons. Then, they can be treated as separate
  contributions in such a way that each nucleon propagator in the RPA ring diagrams is
  modified by multiple-scattering effects before performing the RPA
  summation. In this section, we discuss how the RPA correlation affects the
  structure function as well as the related rates.

\begin{figure*}[htbp]    
%\epsscale{0.87}  
\plotone{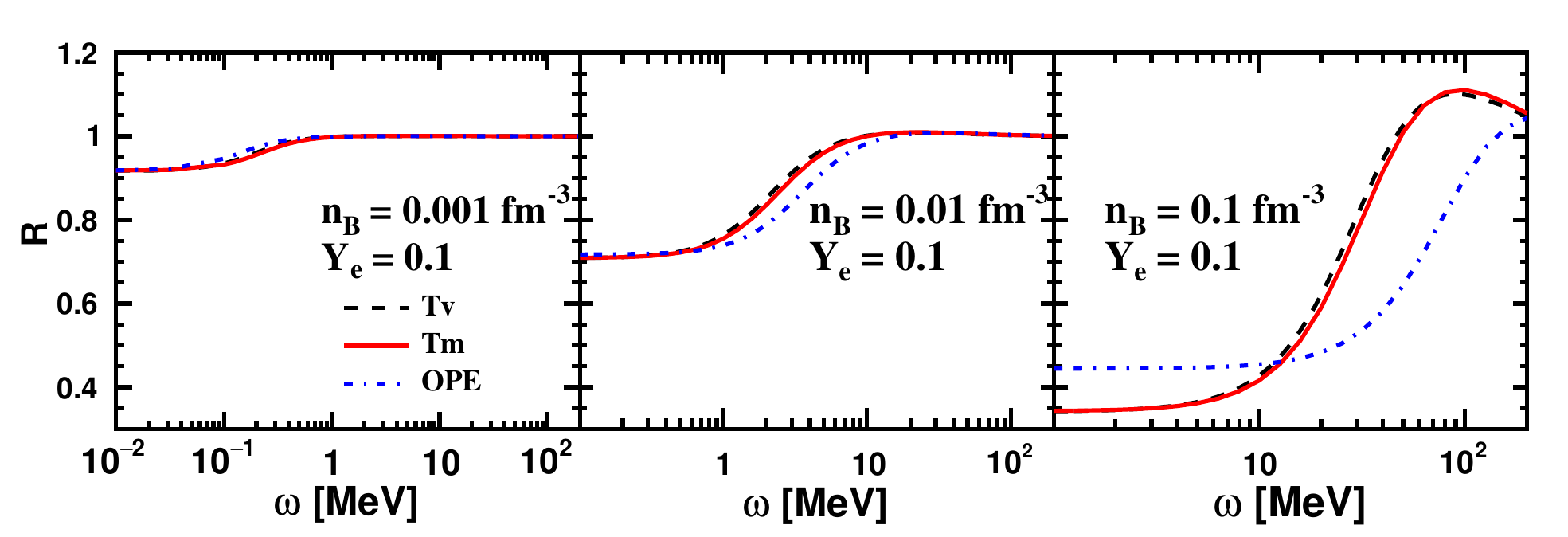}    
\caption{Effects of the RPA correlation on $S_\sigma(\omega)$ at $n_B=0.001$ fm$^{-3}$, 0.01 fm$^{-3}$, and 0.1 fm$^{-3}$, respectively, for $Y_e = 0.1$. $R\equiv S_\sigma^\text{RPA}(\omega)/S_\sigma(\omega)$.}            
\label{fig:SRPA_omega}  
\end{figure*}

The RPA provides a formalism to account for the correlation effects by
summing an infinite number of ring diagrams
\citep{Fetter:1971,Burrows.Sawyer:1998,Reddy.Prakash.ea:1999}. Taking
number density correlation for systems composed of one species as an
example, the polarization function at the RPA level takes the form
$\Pi_\text{RPA}(q, \omega)=\Pi_u(q, \omega)/[1-v(q)\Pi_u(q, \omega)]$,
where $\Pi_u(q, \omega)$ is the polarization function without the RPA correlation and $v(q)$ is the
spin-independent potential. As adopted
in previous literature
\citep{Burrows.Sawyer:1998,Reddy.Prakash.ea:1999}, $\Pi_u(q, \omega)$
can be taken to be the free polarization function
$\Pi^{(0)}(q, \omega)$, which has an analytical expression and is the
same for both density and spin-density
correlations
\citep{Burrows.Sawyer:1998,Reddy.Prakash.Lattimer:1998}. In this work, we choose to consider the RPA
corrections on top of our calculated structure function
$S_\sigma(\omega)$, which already incorporates the multiple-scattering effects.  We follow
the formalism of~\citet{Burrows.Sawyer:1998} based on a spin-dependent
potential~\citep[see
also][]{Reddy.Prakash.ea:1999,Horowitz.Schwenk:2006,
  Horowitz.Caballero.ea:2017}. In principle, RPA calculations should
be based on the same chiral potential as the one used for the
$T$-matrix~\citep{Entem.Machleidt.Nosyk:2017}. However, the choice
of~\citet{Burrows.Sawyer:1998} leads to nucleon scattering rates
within $\sim$ 10\% of the model-independent studies based on virial
expansion in the low-density
region~\citep{Horowitz.Schwenk:2006,Horowitz.Caballero.ea:2017}. On
general grounds the effects of the RPA on the bremsstrahlung rates are expected
to be smaller than for scattering; hence following the approach
of~\citet{Burrows.Sawyer:1998} provides a simple-to-implement method
to quantify their relevance. 

\begin{figure*}[htbp]   
%\epsscale{0.87}  
\plotone{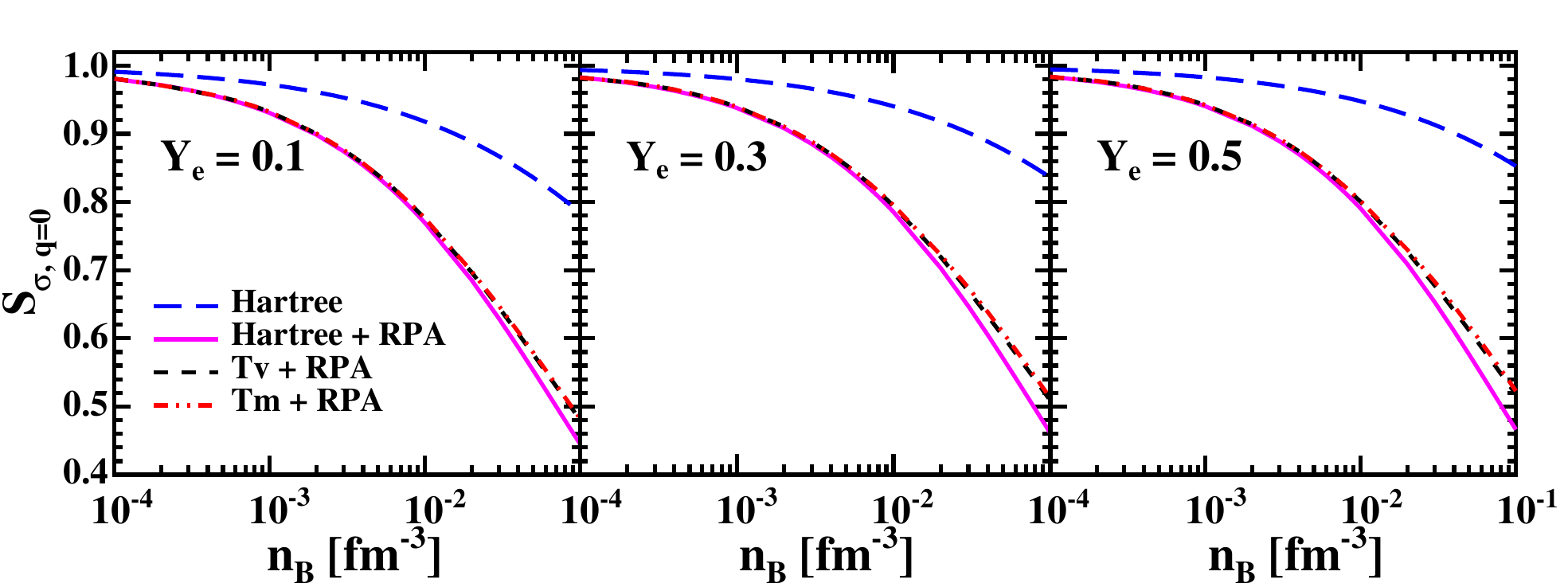}   
\caption{Static structure function in the long-wavelength limit, $S_{\sigma, q=0}$, as functions of density for $Y_e=0.1$, 0.3, and 0.5, respectively.}            
\label{fig:S_q0}  
\end{figure*} 

For a nuclear system consisting of protons and neutrons, the
axial structure function (or the corresponding polarization functions)
takes a $2\times 2$ matrix form as~\citep{Burrows.Sawyer:1998}
\begin{align}
\hat S_\sigma = \left[ \begin{array}{cc} S_\sigma^{(pp)} & \frac{1}{2}S_\sigma^{(pn)}  \\ \frac{1}{2}S_\sigma^{(np)} & S_\sigma^{(nn)} \end{array} \right],     
\end{align}
where the different entries in the matrix contain contributions due to
NN collisions described by the $T$-matrix. The total axial structure
function $S_\sigma$ entering is given by
$S_\sigma =
S_\sigma^{(pp)}+S_\sigma^{(nn)}+\frac{1}{2}(S_\sigma^{(pn)}+S_\sigma^{(np)})=S_\sigma^{(pp)}+S_\sigma^{(nn)}+S_\sigma^{(np)}$,
see Equation~\eqref{eq:Stot}. Despite the fact 
that~\citet{Burrows.Sawyer:1998} consider scattering and we are
interested in bremsstrahlung, we find that their Equation~(47) also applies
to our case, and the structure function that includes both
collision effects based on the $T$-matrix and the RPA correlation is
given by
\begin{align}
S_\sigma^\text{RPA}(\omega) = \frac{2}{n_B} \text{Im}[\Pi(\omega)][1-\exp(-\omega/T)]^{-1} {\cal C}_A^{-1}, 
\end{align}    
where 
\begin{align}
{\cal C}_A = \big\{1-v_\text{GT}\text{Re}[\Pi(\omega)]\big\}^2 + v_\text{GT}^2 \text{Im}[\Pi(\omega)]^2,  
\end{align} 
with $v_\text{GT} = 4.5\times 10^{-5}~\text{MeV}^{-2}$ and  
$\Pi(\omega)$ is given by       
\begin{align}
\text{Im}[\Pi(\omega)] =& {n_B\over 2}S_\sigma(\omega) [ 1-\exp(-\omega/T)],  \\ 
\text{Re}[\Pi(\omega)] =& {1\over \pi} {\cal P}  \int d\omega'  {\text{Im}[\Pi(\omega')] \over \omega-\omega'}.  
\end{align}

Figure~\ref{fig:SRPA_omega} shows how the normalized structure functions in Equation~\eqref{eq:norm} based on different nuclear
matrix elements are affected by RPA in different
conditions with
$R\equiv S^\text{RPA}_\sigma(\omega)/S_\sigma(\omega)$. The effect of
the RPA correlation is to reduce $S_\sigma(\omega)$ at low $\omega$ due to a negative $\text{Re}[\Pi(\omega)]$, and to increase it
slightly at high $\omega$ as $\text{Re}[\Pi(\omega)]$ turns
positive. We also show in Figure~\ref{fig:S_q0} the effects of the RPA on the
static structure function [or the normalization of $S_\sigma(\omega)$;
see Equation~(\ref{eq:norm})] in the long-wavelength limit, which is
defined as
\begin{equation}
S_{\sigma, q=0} \equiv \lim_{q\to 0} \int_{-\infty}^{\infty} \frac{S_\sigma(q, \omega)}{2\pi} d\omega 
= \int_{-\infty}^{\infty} \frac{S_\sigma(\omega)}{2\pi} d\omega.
\end{equation} 
For comparison, we also present the mean-field or Hartree 
result~\citep{Reddy.Prakash.Lattimer:1998}, which is simply given
by Equation~(\ref{eq:norm}), the same as the static structure function associated with our normalized $S_\sigma(\omega)$ without
including the effects of the RPA. The RPA correlations reduce $S_{\sigma, q=0}$,
consistent with the studies based on virial
expansion~\citep{Horowitz.Schwenk:2006,Horowitz.Caballero.ea:2017}. Furthermore, we
find a very similar reduction due to RPA correlations for the
mean-field case and for cases that consider collisions based on the
$T$-matrix. This justifies our assumption that the nucleon-nucleon
collisional broadening does not affect the normalization of
$S_\sigma(\omega)$, but just redistributes the strength in a broader
energy region.

\begin{figure}[htbp] 
\centering  
%\epsscale{1.0}  
\includegraphics[width=\linewidth]{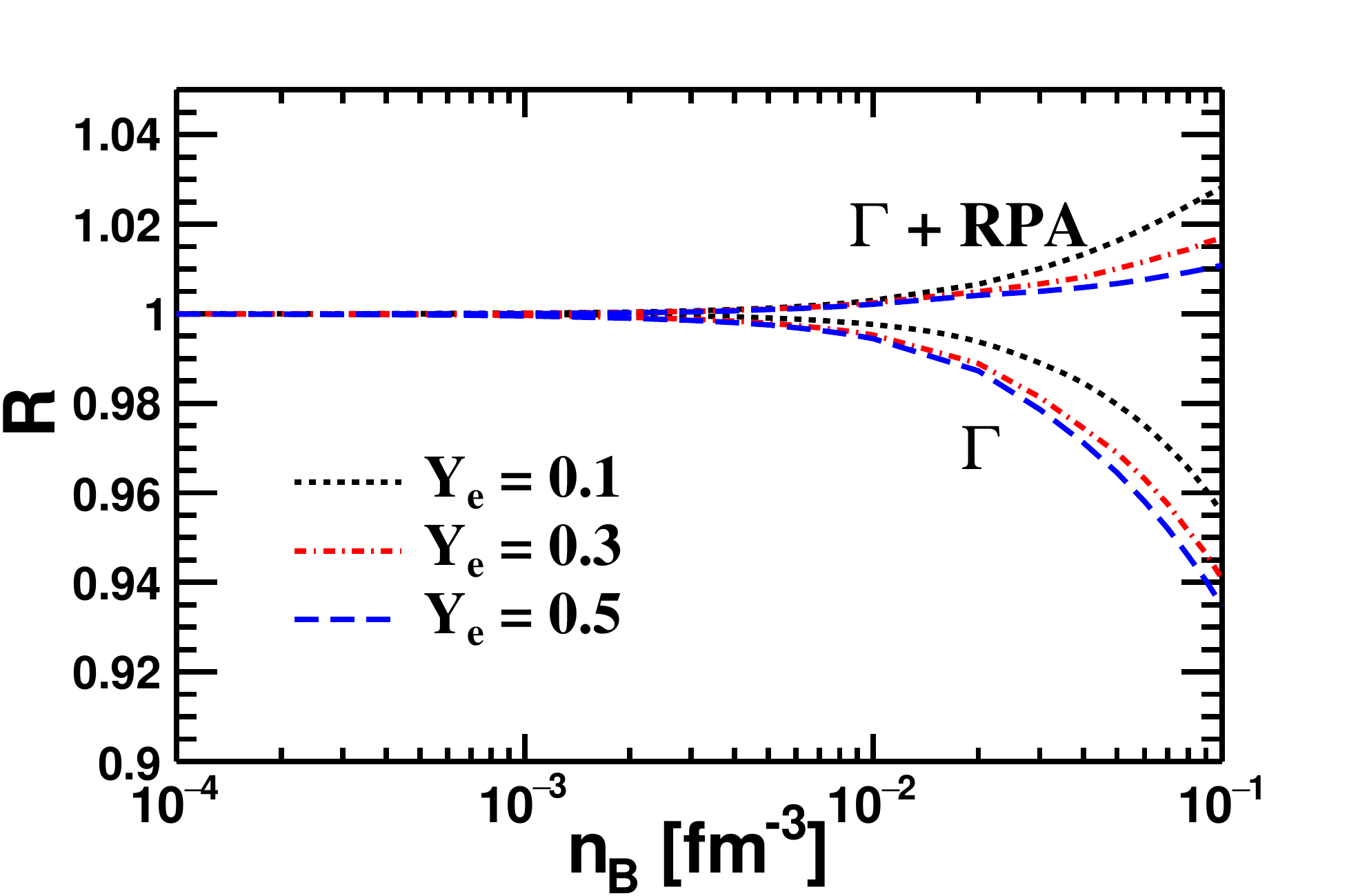}
\caption{Ratios of $\langle \lambda_A^{-1} \rangle$ considering the
  effects of $\Gamma$ and the RPA correlation to those based on the
  unnormalized structure functions, as functions of density for
  $Y_e$=0.1, 0.3, and 0.5, respectively.\label{fig:opacity_rpa}}
\end{figure}   

\begin{figure}[htbp] 
\centering  
%\epsscale{1.0}  
\includegraphics[width=\linewidth]{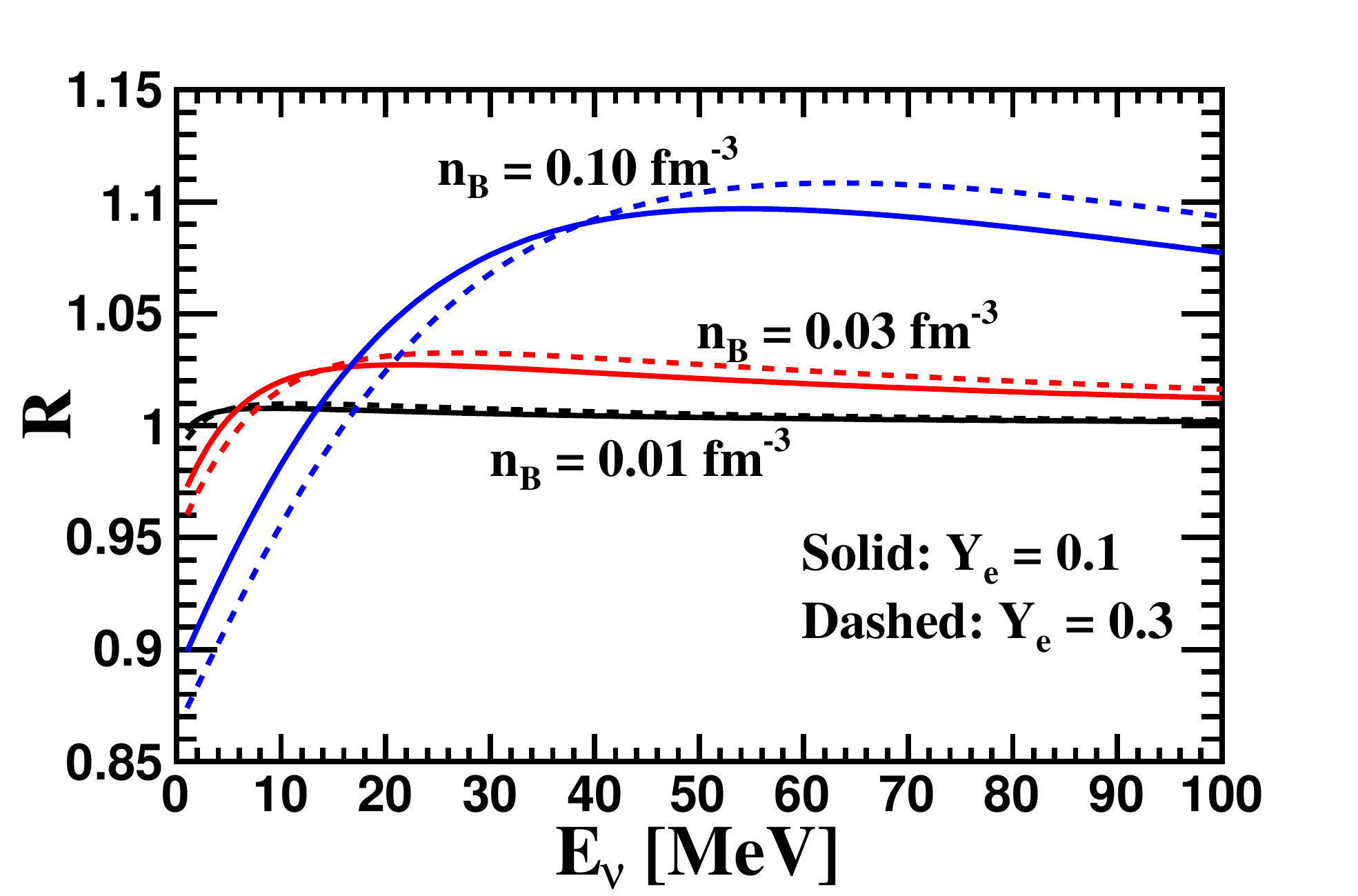}          
\caption{Ratios of $\lambda_A^{-1}(E_\nu)$ with the RPA effect to
those without at $n_B=0.01$, 0.03, and 0.1 fm$^{-3}$ and $Y_e$=0.1 and 0.3,
respectively. The in-medium $T$-matrix elements are used for
calculating $S_\sigma(\omega)$.\label{fig:ratio_rpa_e}}
\end{figure}

Figure~\ref{fig:opacity_rpa} shows the effects of the RPA correlations
and the width parameter $\Gamma$ on $\langle \lambda_A^{-1} \rangle$
as a function of density. As discussed above, the inclusion of $\Gamma$
only affects $S_\sigma(\omega)$ for $\omega \lesssim \Gamma$. However,
$\langle \lambda_A^{-1} \rangle$ is determined by $S_\sigma(\omega)$
at high $\omega$ and hence the effect of $\Gamma$ is rather
insignificant, reducing the rates by up to a few per cent at
subsaturation densities. The average rate,
$\langle \lambda_A^{-1} \rangle$, is enhanced slightly
by the effects of the RPA, due to the increased $S_\sigma(\omega)$ in the high-$\omega$
region; see Figure~\ref{fig:SRPA_omega}. The combined effect of
$\Gamma$ and RPA is $\sim$ 3\% at most.

The effect of the RPA correlations on the energy-dependent inverse
mean free path $\lambda_A^{-1}(E_\nu)$ is illustrated in
Figure~\ref{fig:ratio_rpa_e} by showing the ratio of
$\lambda^{-1}_A(E_\nu)$ including RPA correlations to that
without. The impact is similar to that on $S_\sigma(\omega)$ shown
in Figure~\ref{fig:SRPA_omega}, i.e., a suppression in the low-energy 
region and an enhancement at high energy. We find that the effects
become significant only for $n_B\gtrsim 0.01~\text{fm}^{-3}$ and can
reach up to 10\% near the saturation density, consistent with the results
shown in Figure~\ref{fig:opacity_rpa} for
$\langle \lambda_A^{-1} \rangle$.

\section{summary and conclusions}
\label{sec:sum}

We have revisited the rate of neutrino bremsstrahlung in supernova
matter in the long-wavelength limit and investigated the effects of
different treatments in a systematic way. The vacuum/in-medium
$T$-matrix for NN scattering with half-off-shell elements obtained by
solving the Lippmann-Schwinger/Bethe-Goldstone equation based on
$\chi$EFT potentials has been used to study the bremsstrahlung rates,
to be compared with those based on the OPE potential and the
associated diagonal/on-shell matrix elements. For a broad range of
density, temperature, and $Y_e$ relevant to supernova conditions, we
have considered the blocking of the final nucleons, which is to be
compared with the studies using the Boltzmann distribution without
blocking. We have also explored the effects of the width parameter, to
account for multiple-scattering effects, and the RPA correlations on the
structure function and the related rates. A numerical table of our new structure function based on the vacuum $T$-matrix is provided (see more details in Appendix~\ref{sec:table}).

Taking Equation~(\ref{eq:SN_T}) to characterize the typical SN conditions, 
our studies show that ignoring the blocking of the final nucleons
overestimates the rates by $\sim$ 20\% at $n_B=0.01$ fm$^{-3}$
($\rho \approx 1.7\times 10^{13}$~g~cm$^{-3}$) and by $\sim$
50\%--100\% at $n_B=0.1$ fm$^{-3}$
($\rho \approx 1.7\times 10^{14}$~g~cm$^{-3}$). Using the effective
diagonal/on-shell $T$-matrix elements underestimates the rates by
$\sim$ 50\%--70\% at $n_B=10^{-4}$~fm$^{-3}$
($\rho \approx 1.7\times 10^{11}$~g~cm$^{-3}$), and by $\sim$
30\%--50\% at $n_B=10^{-3}$~fm$^{-3}$
($\rho \approx 1.7\times 10^{12}$~g~cm$^{-3}$), with the effects
getting stronger with increasing $Y_e$. Close to the saturation
density, the effective on-shell $T$-matrix gives rise to an
enhancement by $\sim$ 20\%--40\%. We therefore argue that the
half-off-shell $T$-matrix elements are required for an accurate study
of the bremsstrahlung rate. We confirm the results of previous
studies~\citep{Bartl.Pethick.Schwenk:2014,Bartl:2016} that using the
$T$-matrix element instead of the OPE potential leads to an
enhancement by a factor of 2--5 at $n_B=10^{-4}$~fm$^{-3}$
($\rho \approx 1.7\times 10^{11}$~g~cm$^{-3}$) due to the resonant
property of the NN force, and a suppression at densities above
$\sim 2\times 10^{-3}$ fm$^{-3}$
($\rho \approx 3.3\times 10^{11}$~g~cm$^{-3}$). For the supernova-relevant conditions explored in
this paper [see Equation~\eqref{eq:SN_T}], we find that the results
obtained using the standard vacuum $T$-matrix are very similar to those
based on the in-medium $T$-matrix. Nevertheless, we expect that the
differences will be larger for
cold neutron stars where nucleons are highly degenerate. 

Following~\citet{Hannestad.Raffelt:1998}, we introduce a width
parameter or spin relaxation rate $\Gamma$ to normalise the axial
structure function $S_\sigma(\omega)$ and to make a proper comparison
with the previous studies in the
literature~\citep{Raffelt:2001,Lykasov.Pethick.Schwenk:2008,Bacca.Hally.ea:2009,Bacca.Hally.ea:2012,Bartl.Pethick.Schwenk:2014}. The
effect of $\Gamma$ is to suppress $S_\sigma(\omega)$ at low $\omega$, and we find that the rates based on the normalized
$S_\sigma(\omega)$ are only reduced by a few per cent for densities
above $\sim$ 0.01 fm$^{-3}$
($\rho \approx 1.7\times 10^{13}$~g~cm$^{-3}$). Comparisons of
neutrino pair absorption/emission rates based on our normalized
$S_\sigma(\omega)$ to those from
\cite{Hannestad.Raffelt:1998} and \citet{Bartl.Pethick.Schwenk:2014} are
summarized in Figures~\ref{fig:Opa_absorption} and \ref{fig:emis_E}. We find that the relative ratios of our results using the
$T$-matrix to those from the previous literature could be either as
small as $\sim$ 0.2, or as large as $\sim$ 5 for different regions of
density and $E_\nu$ considered. The difference from
\citet{Bartl.Pethick.Schwenk:2014} originates mainly from off-shell
effects on the $T$-matrix as well as the blocking of the final
nucleons.

Effects of the RPA correlation on top of the normalised
$S_\sigma(\omega)$ that incorporates collisional broadening are
further explored. We find that $S_\sigma(\omega)$ is reduced significantly
at low $\omega$ and slightly enhanced at high $\omega$. Though the
normalization of $S_\sigma(\omega)$ is reduced, which is consistent with the prediction from virial expansion, the energy-averaged
inverse mean free path $\langle \lambda_A^{-1} \rangle$ is slightly
enhanced by $\lesssim$ (2--3)\% below the saturation density. Similarly to
$S_\sigma(\omega)$, $\lambda_A^{-1}(E_\nu)$ is suppressed at low
$E_\nu$ and enhanced at high $E_\nu$ by the RPA correlations, but only
by a negligible factor that is within a few per cent for the relevant conditions.

The impact of neutrino bremsstrahlung rates beyond OPE has been
explored in 1D supernova
simulations~\citep{Bartl.Bollig.ea:2016,Fischer:2016}~\citep[see
also][for studies based on bremsstrahlung rates using the OPE
potential]{Raffelt:2001,Keil.Raffelt.Janka:2003}.
Our calculations based on the half-off-shell $T$-matrix, similarly to
those of~\citet{Bartl.Pethick.Schwenk:2014}, predict a low-density
resonant enhancement of the bremsstrahlung rate
$\langle \lambda_A^{-1} \rangle$ and a suppression at high densities
when compared with the OPE results (see Figure~\ref{fig:Opa_absorption}).
\citet{Bartl.Pethick.Schwenk:2014} predict a transition density that
is typically smaller than the value reached at the neutrinosphere,
which moves from $\rho \sim 10^{12}$~g~cm$^{-3}$ to
$10^{14}$~g~cm$^{-3}$ as the PNS deleptonizes. Hence,
the net effect in the supernova simulations
of~\citet{Bartl.Bollig.ea:2016} is a reduction of the bremsstrahlung
rate by a factor of $\sim$ 2--5 when compared with the OPE rates. Such
a reduction translates into a minor change in the neutrino luminosities
($\lesssim 5\%$) and a small increase of the averaged neutrino energies, $\langle E_\nu \rangle$, within 1~MeV. In our calculations, the transition density is shifted
to higher densities similar to those in the neutrinosphere. This
may indicate an even smaller impact on the neutrino emission of the
rates presented here. However, given the nonlinear nature of neutrino
processes in supernova matter, a fully self-consistent simulation is
required to quantify their impact.

We expect that the improved treatment of the NN interaction presented in
this work will significantly affect the inelastic scattering process,
$\nu+N+N\to \nu+N+N$, which should exhibit more relevance in SN
dynamics~\citep{Sawyer:1995,Raffelt.Strobel:1997,Hannestad.Raffelt:1998,Raffelt:2001,Melson.Janka.ea:2015,Burrows.Vartanyan.ea:2018,Kotake.Takiwaki.ea:2018}. In
principle, the same structure function, $S_\sigma(q, \omega)$, governs
both neutrino scattering and bremsstrahlung as well as pair absorption
in the nuclear medium, though different regions in $(q, \omega)$ are
relevant to each process, i.e., $q^2 \le \omega^2$ for pair absorption
and bremsstrahlung, and $q^2\ge \omega^2$ for inelastic scattering. To
obtain $S_\sigma$ in this work, we have taken the long-wavelength
limit and therefore ignored the recoil of nucleons. This is a good
approximation for studying pair absorption and bremsstrahlung, since
the recoil energy, $E_r$, is always negligible compared to the energy
transfer or the width parameter, i.e.,
$E_r \sim q P_N/m_N \sim (T/m_N)^{1/2}q \lesssim (T/m_N)^{1/2}\omega
\ll \text{max}(\omega, \Gamma)$ with $P_N$ the typical nucleon
momentum. We have checked by using the OPE potential that nucleon
recoil affects the associated rates by only a few per cent for typical
SN conditions. However, this is not the case for inelastic scattering,
since $q$ is typically of the order of $E_\nu$ or $E_\nu'$, and could be
much larger than $\omega=E_\nu-E_\nu'$, which vanishes in the elastic
limit. Therefore, the nucleon recoil is no longer negligible
\citep{Raffelt:2001}. We argue that the studies of $S_\sigma(\omega)$
in this work should still be reliable for studying neutrino scattering
in the limit of $E_r \sim \text{max}(E_\nu, E_\nu') \times (T/m_N)^{1/2} \ll
\Gamma$. In the opposite limit where $\Gamma$ is negligible compared
to $E_r$ and $\omega$, $S(q, \omega)$ has an analytical expression
including the recoil
effect~\citep{Burrows.Sawyer:1998,Reddy.Prakash.Lattimer:1998,Raffelt:2001}. We
plan to provide a full treatment of $S_\sigma(q, \omega)$
incorporating both the collisional broadening and the nucleon recoils
in the future work.
      
\acknowledgments We gratefully acknowledge David Rodr\'iguez Entem for
providing us with the Fortran code of the NN $\chi$EFT potential. We
thank Bengt Friman, Thomas Neff, Sanjay Reddy, Achim Schwenk,
Stefan Typel, and Hans-Thomas Janka for stimulating discussions about this project. We thank Aurore Betranhandy for testing our rates in supernova simulations and helping to correct a mistake in Figure~8. We also thank the anonymous referee for reading the paper carefully and providing helpful comments. This work
has been partly supported by the Deutsche Forschungsgemeinschaft (DFG,
German Research Foundation) - Projektnummer 279384907 - SFB 1245.
      
\appendix

\section{$T$-matrix elements in vacuum/nuclear medium} 
\label{sec:app_tmatrix}
   
The relevant formulae for obtaining the NN scattering $T$-matrix elements in vacuum/nuclear medium are shown below.     
\subsection{vacuum $T$-matrix}  
The vacuum $T$-matrix can be obtained from the Lippmann-Schwinger (LS) equation as 
\begin{align}
\mathcal{T}(\bm{k}', \bm{k}; E) = V(\bm{k}', \bm{k}) + \int {d^3k'' \over (2\pi)^3} V(\bm{k}', \bm{k}'') {1 \over E-{k''^2\over m_N} + i\varepsilon} \mathcal{T}(\bm{k}'', \bm{k}; E),  
\label{eq:LS}
\end{align}
where $\bm{k}$ ($\bm{k}')$ is the relative momentum between the two
incoming (outgoing) nucleons, and we adopt the notation
$k\equiv |\bm{k}|$. We usually call the $T$-matrix on-shell when
$E=k'^2/m_N=k^2/m_N$, half-off-shell when one of them holds, or
off-shell when neither holds.

In partial wave components, the LS equation can be cast as 
\begin{align}
\mathcal{T}^{JST}_{ll'}(k', k; E) = V^{JST}_{ll'}(k', k) + \sum_{l''}\int {k''^2dk'' \over (2\pi)^3} V^{JST}_{ll''}(k', k'') {1 \over E-{k''^2\over m_N} + i\varepsilon} \mathcal{T}^{JST}_{ll''}(k'', k; E), 
\label{eq:LS_pw}
\end{align} 
where indices $J, S$ and $T$ are the three conserved quantum numbers:
the total angular momentum, the total spin, and the total isospin of
the nucleon pair; and $l, l'$ are the relative orbital angular
momenta for the incoming and outgoing nucleon pairs. Note that the
coupling of partial waves with different $l$ arises from the tensor
part of the nuclear force, which does not conserve the angular momentum
$l$. Due to the conservation of $J$ ($\vec J=\vec l+\vec S$) and
parity $\Pi = (-1)^l$, the only allowed values of
$\Delta l \equiv l-l'$ are 0, $\pm 2$. Another selection rule from the
Pauli exclusion principle is that $l+S+T$ should be odd.

The LS equation [Equation~(\ref{eq:LS_pw})] can be numerically solved by matrix inversion after discretizing the integral into a sum \citep{Haftel.Tabakin:1970,Machleidt:1993,Machleidt:2001}. Note that the factor $i\varepsilon$ in the denominator coupling the real and imaginary parts of the $T$-matrix makes the calculation more involved.\footnote{A direct solution for the complex $T$-matrix from matrix inversion is also carried out, which can provide a crosscheck for the $R$-matrix calculations. It has been shown that high-precision consistency between these two approaches can always be reached.} A more efficient way is to deal with the real $R$-matrix (or $K$-matrix), which is defined as \citep{Landau:1990} 
\begin{align} 
\mathcal{R} = \mathcal{T} + i\pi \mathcal{T} \delta(E-H_0)\mathcal{R}, \label{eq:R-T}
\end{align}
and the $R$-matrix obeys a `real' version of the LS equation as  
\begin{align}
\mathcal{R}^{JST}_{ll'}(k', k; E) = V^{JST}_{ll'}(k', k) + \sum_{l''}\int {k''^2dk'' \over (2\pi)^3} V^{JST}_{ll''}(k', k'') {\cal P}\Big[{1 \over E-{k''^2\over m_N}}\Big] \mathcal{R}^{JST}_{ll''}(k'', k; E).   
\label{eq:LSr_pw}
\end{align}            

Once $\mathcal{R}^{JST}_{ll'}$ is known, we can obtain the components of the $T$-matrix
from Equation~(\ref{eq:R-T}). For a half-off-shell $T$-matrix, we need to
solve
\begin{align}
\mathcal{T}^{JST}_{ll'}(k, k_0; E_{k_0}) \Big[\delta_{l'l''} + i\pi
  \frac{m_N k_0}{2(2\pi)^3}\mathcal{R}^{JST}_{l'l''}(k_0, k_0;
  E_{k_0})\Big] = \mathcal{R}^{JST}_{ll''}(k, k_0; E_{k_0}) ,  
\label{eq:R-T2}    
\end{align}   

with $E_{k_0}=k_0^2/m_N$. This can be further simplified for uncoupled channels (with $l=l'$) to
\begin{align}
\mathcal{T}^{JST}_{ll}(k, k_0; E_{k_0}) = { \mathcal{R}^{JST}_{ll}(k, k_0; E_{k_0}) \over 1 + i\pi {m_N k_0 \over 2(2\pi)^3} \mathcal{R}_{ll}(k_0, k_0; E_{k_0}) }, 
\label{eq:R-T3_unc}  
\end{align} 
with 
\begin{align}
& \text{Re}[\mathcal{T}^{JST}_{ll}(k, k_0; E_{k_0})] = { \mathcal{R}^{JST}_{ll}(k, k_0; E_{k_0}) \over 1 + [\pi {m_N k_0 \over 2(2\pi)^3} \mathcal{R}^{JST}_{ll}(k_0, k_0; E_{k_0})]^2 },  \\
& \text{Im}[\mathcal{T}^{JST}_{ll}(k, k_0; E_{k_0})] = { -\pi {m_N k_0 \over 2(2\pi)^3} \mathcal{R}^{JST}_{ll}(k, k_0; E_{k_0}) \mathcal{R}^{JST}_{ll}(k_0, k_0; E_{k_0})  \over 1 + [\pi {m_N k_0 \over 2(2\pi)^3} \mathcal{R}^{JST}_{ll}(k_0, k_0; E_{k_0})]^2 }. 
\end{align} 

The phase shifts for uncoupled channels are simply given by \citep{Machleidt:2001}
\begin{align}
\tan\delta^{JST}_{l}(E_{k_0})= -\frac{\pi k_0 m_N }{2(2\pi)^3} \mathcal{R}^{JST}_{ll}(k_0, k_0; E_{k_0}),            
\end{align}
and for coupled channels we have 
\begin{align}
U^{-1}\left[ \begin{array}{cc} \tan\delta^{J}_{J-1}(E_{k_0})  & 0 \\ 0
                                                              &
                                                                \tan\delta^{J}_{J+1}(E_{k_0})  \end{array} \right]U = -\frac{\pi k_0 m_N}{2(2\pi)^3} \left[ \begin{array}{cc}  \mathcal{R}^{J}_{J-1,J-1}(k_0, k_0; E_{k_0}) & \mathcal{R}^{J}_{J-1,J+1}(k_0, k_0; E_{k_0})  \\ \mathcal{R}^{J}_{J+1,J-1}(k_0, k_0; E_{k_0}) & \mathcal{R}^{J}_{J+1,J+1}(k_0, k_0; E_{k_0}) \end{array} \right],   
\label{eq:coupled}
\end{align}
with $U$ the standard $2 \times 2$ mixing matrix given by  
\begin{align}
U = \left[ \begin{array}{cc} \cos\varepsilon^J & \sin\varepsilon^J \\ -\sin\varepsilon^J  & \cos\varepsilon^J \end{array} \right].   
\end{align}
Note that we are allowed to drop the indices $S, T$ for coupled channels without introducing any confusion since they are fixed for a given $J$ (i.e., use $\delta^J_{J\pm 1}$, $\varepsilon^J$ and $\mathcal{R}^J_{J\pm 1, J\pm 1}$). From Equation~(\ref{eq:coupled}), we can obtain 
\begin{align}
& \tan2\varepsilon^J(E_{k_0}) = \frac{2 \mathcal{R}^{J}_{J+1,J-1}} {\mathcal{R}^{J}_{J-1,J-1}-\mathcal{R}^{J}_{J+1,J+1}}, \\ 
& \tan\delta^J_{J\mp 1}(E_{k_0}) = -\frac{\pi  k_0 m_N}{4(2\pi)^3} \Big[\mathcal{R}^{J}_{J-1,J-1} + \mathcal{R}^{J}_{J+1,J+1} \pm \frac{ \mathcal{R}^{J}_{J-1,J-1} - \mathcal{R}^{J}_{J+1,J+1}  }{\cos2\varepsilon^J(E_{k_0})}\Big]. \label{eq:delta_c}       
\end{align}

Once the phase parameters $\varepsilon^J$ and $\delta^J_{J\pm 1}$ are fixed from the $R$-matrix elements, the $T$-matrix elements for the coupled channels can be obtained from Equation~(\ref{eq:R-T2}) as
\begin{align}
\mathcal{T}^J_{ll'}(k, k_0; E_{k_0}) = \mathcal{R}^J_{ll''}(k, k_0; E_{k_0}) U^{-1}_{l'' \tilde l}  \left[ \begin{array}{cc}  1-i \tan\delta^J_{J-1} & 0 \\ 0 & 1-i \tan\delta^J_{J+1}  \end{array} \right]^{-1}_{\tilde l \tilde{\tilde l}} U_{\tilde{\tilde l}l'}.   
\label{eq:R-T3_c}  
\end{align}   
    
The above phase shifts and mixing parameters for coupled channels are
defined in the so-called `BB' convention
\citep{Blatt.Biedenharn:1952}. An alternative convention for the phase
parameters is proposed in \citet{Stapp.Ypsilantis.Metropolis:1957},
which is known as the `bar' convention, and is usually adopted for
analyzing NN scattering data. The two conventions are the same for
uncoupled channels but different for the coupled channels. In the
`bar' convention, the on-shell $T$-matrix is given by
\begin{align}
\mathcal{T}^{JST}_{ll'}(k_0, k_0; E_{k_0}) = -\frac{8\pi^2}{im_N k_0} [\exp(2i \bar\delta^{JST}_{l}) -1]   \label{eq:phase_bar}
\end{align} 
for uncoupled channels, and 
\begin{align}
\mathcal{T}^{JST}_{ll'}(k_0, k_0; E_{k_0}) = -\frac{8\pi^2}{im_N k_0} \times \left\{ \begin{array}{c} [\exp(2i\bar\delta^{J}_{l}) \cos2\bar\varepsilon^J -1] ~~\text{for $l=l'$,} \\ i\exp[i(\bar\delta^{J}_{l} + \bar\delta^J_{l'})] \sin2\bar\varepsilon^J ~~\text{for $l \ne l'$,} \end{array} \right.  \label{eq:phase_bar_uncoup}        
\end{align}   
for coupled channels. Note that only the on-shell $T$-matrix elements
can be obtained from measured phase shifts; in order to obtain the
off-shell elements, the LS equation should be solved based on a given nuclear potential.    

\subsection{in-medium $T$-matrix}

The discussion for the vacuum $T$-matrix can be applied to the in-medium $T$-matrix using the Bethe-Goldstone (BG) equation,  
\begin{align}
\mathcal{T}^{JST}_{ll'}(k', k; K, \Omega) = V^{JST}_{ll'}(k', k) + \sum_{l''}\int {k''^2dk''\over (2\pi)^3} V^{JST}_{ll''}(k', k'') \bar {g}_{II}(K, \Omega, k'') \mathcal{T}^{JST}_{l''l}(k'', k; K, \Omega), 
\label{eq:LS_m}
\end{align}      
where $\bar {g}_{II}(K, \Omega, k'')$ is an angle-averaged two-particle propagator,\footnote{the angle-averaged procedure is applied to avoid coupling of partial waves with different values of $J$; and only minor effects are introduced compared to the exact procedure with the full two-particle propagator \citep{Sartor:1996,Suzuki.Okamoto.ea:2000,Frick.Gad.ea:2002}.} which in the quasiparticle approximation is given by    
\begin{align}
\bar {g}_{II}(K, \Omega, k) = \biggl\langle \frac{1-f(\varepsilon(k_1))-f(\varepsilon(k_2))}{\Omega - \varepsilon(k_1) - \varepsilon(k_2)+ i \eta} \biggr\rangle_{\theta},    
\label{eq:gII}  
\end{align}
with the two nucleon momenta
$\bm{k}_{1,2} = {1\over 2}\bm{K} \pm \bm{k}$ and $\theta$ the angle
between the total momentum $\bm{K}$ and the relative momentum
$\bm{k}$. $\Omega$ is the total energy of the nucleon pair and
$\varepsilon(k_{1,2})$ is the non-relativistic single-particle energy
of the nucleon that in the quasiparticle approximation can be given by
$\varepsilon(k_{i})=k_{i}^2/(2m^*_{n,p})+U_{n,p}$, where $m^*_{n,p}$ and $U_{n,p}$ are the effective masses and interacting potentials of neutron and proton in the nuclear medium. The Fermi function takes
the standard form as
$f(\varepsilon) = 1/[ 1 + \exp\big(\frac{\varepsilon -
  \mu}{k_BT}\big)]$ with $\mu$ the non-relativistic chemical potential
of the nucleon. In the low-density limit, we have
$\varepsilon(k_{1,2}) \to \varepsilon_0(k_{1,2})=k_{1,2}^2/(2m_N)$ and $1-\varepsilon((k_1))-f(\varepsilon(k_2))\to 1$, and therefore
$\bar {g}_{II}(K, \Omega, p) \to (E-k^2/m_N+i\eta)^{-1}$ with
$E = \Omega - K^2/(4m_N)$ and $m_N=(m_n+m_p)/2$ the averaged nucleon bare mass. 

Exactly the same procedures are taken to numerically solve the
in-medium $T$-matrix. Once the partial wave components of the $R$-matrix are
obtained from matrix inversion, one can use Equations~(\ref{eq:R-T3_unc})
and (\ref{eq:R-T3_c}) to obtain the $T$-matrix elements, but with the
replacement of $k_0m_N$ by
$2k_0^2 \langle 1-f(\varepsilon(k_1))-f(\varepsilon(k_2))
\rangle_\theta [d\langle \varepsilon(k_1) + \varepsilon(k_2)
\rangle_\theta/dk|_{k=k_0}]^{-1}$ at
$\Omega = \langle\varepsilon(k_1) + \varepsilon(k_2)\rangle_\theta$
for any given value of $K$ [see Equations~(\ref{eq:R-T3_unc})
and (\ref{eq:delta_c})].  In the low-density limit we have
$\langle \varepsilon(k_1) + \varepsilon(k_2) \rangle_\theta = {K^2
  \over 4m_N} + {k^2 \over m_N}$, $f(\epsilon(k_{1,2})) \ll 1$, and
therefore
$d\langle \varepsilon(k_1) + \varepsilon(k_2)
\rangle_\theta/dk|_{k=k_0} = 2k_0/m_N$, which guarantees that the
in-medium $T$-matrix goes to the vacuum one.

Throughout this work, we always take the bare nucleon mass for all our studies. For bremsstrahlung, the nucleon interaction potentials can be absorbed into the chemical potentials and we can simply take $U_{n,p}$=0 without affecting the final results. As can be easily seen, the medium effects on the $T$-matrix considered in this work are mainly from the blocking factor in Equation~(\ref{eq:gII}).

\section{matrix elements for NN bremsstrahlung in partial wave
  components} 
\label{sec:app_TsqB} 

\citet{Bartl.Pethick.Schwenk:2014} and \citet{Bartl:2016} have
developed a formalism for the calculation of matrix elements of NN 
bremsstrahlung in partial wave components within the long-wavelength
approximation in the $pn$-formalism. In the following, we present an
alternative derivation using the isospin formalism.

Let us consider the process (see Figure~\ref{fig:NNB})
\begin{equation}
  \label{eq:1}
  N_a + N_b \rightarrow N_c + N_d + \nu + \bar{\nu},
\end{equation}
where $N$ stands for either a neutron, $n$, or a proton, $p$. Energy and momentum conservations imply

\begin{subequations}
\begin{eqnarray}
  \label{eq:mom}
  \bm{k}_a + \bm{k}_b & = & \bm{k}_c + \bm{k}_d + \bm{q}, \\
  \label{eq:ener}
  E_a + E_b & = & E_c + E_d + \omega. 
\end{eqnarray}
\end{subequations}
In the following, we will consider the long-wavelength limit in which neutrinos carry
away zero momentum, i.e., $\bm{q} = \bm{0}$. We define the relative
momenta of the nucleons as

\begin{equation}
  \label{eq:2}
  \bm{k}_i = \frac{\bm{k}_a - \bm{k}_b}{2}, \quad \bm{k}_f = \frac{\bm{k}_c - \bm{k}_d}{2}.
\end{equation}
As the center-of-mass momentum of the nucleons is conserved, we
consider states that are characterized by the relative momenta of the
nucleons and their spin projections, $|ab\rangle = |\bm{k}\, s_as_b\rangle$ normalized
such as
\begin{equation}
  \label{eq:3}
  \langle\bm{k}_f\, s_cs_d|\bm{k}_i\, s_as_b\rangle = (2\pi)^3\delta(\bm{k}_i-\bm{k}_f)
  \delta_{s_a s_c} \delta_{s_b s_d}.
\end{equation}
We can build fully antisymmetric states using the permutation operator $P_{12}$ as 

\begin{equation}
  \label{eq:4}
  |ab\rangle_\text{nas} = |\bm{k}\, s_as_b\rangle_{\text{nas}} = \frac{1}{\sqrt{2}} (1-P_{12}) |\bm{k}\,
  s_as_b\rangle = \frac{1}{\sqrt{2}} \Bigl[|\bm{k}\,
  s_as_b\rangle- |-\bm{k}\,s_bs_a\rangle\Bigr].
\end{equation}

States of good spin and isospin are obtained as

\begin{equation}
  \label{eq:5}
  |\bm{k}\,S S_z\, T T_z\rangle = \sum_{s_a s_b t_c t_d} \langle 1/2\, s_a\, 1/2\,
  s_b | S\, S_z \rangle \langle 1/2\, t_a\, 1/2\,
  t_b | T\, T_z \rangle |\bm{k}\, s_as_b\, t_a t_b \rangle,
\end{equation}
where we use the convention of neutrons having isospin projection
1/2. Expanding the plane wave states into partial waves, we obtain

\begin{equation}
  \label{eq:pw}
  |\bm{k}\,S S_z\, T T_z\rangle = \sum_{lm} Y^{*}_{lm}(\hat{\bm{k}})
  |k\,lm \,S S_z\, T T_z \rangle,
\end{equation}
with $\hat{\bm{k}}$ the unit vector in the
direction of $\bm{k}$ and $k=|\bm{k}|$. The partial wave states are
normalized as

\begin{equation}
  \label{eq:6}
  \langle k_f\,l_fm_f\, S_f S_{z,f}\, T_f
  T_{z,f}|k_i\,l_im_i\, S_i S_{z,i}\, T_i
  T_{z,i} \rangle= (2\pi)^3\frac{\delta(k_i-k_f)}{k_i^2} \delta_{l_f l_i}
  \delta_{m_f m_i}\delta_{S_f S_i}\delta_{S_{z,f}
    S_{z,i}}\delta_{T_f T_i}\delta_{T_{z,f}
    T_{z,i}}.
\end{equation}
Coupling the orbital angular momentum and spin, we finally have states

\begin{equation}
  \label{eq:coupleJ}
  |k\, l S J M\, T T_z\rangle = \sum_{m S_z} \langle l\, m\, S\,
  S_z | J\, M \rangle |k\,lm\, S S_z\, T
  T_z\rangle.
\end{equation}
We can build normalized antisymmetric states using the permutation operator, e.g.,

\begin{equation}
  \label{eq:7}
  |\bm{k}\,S S_z\, T T_z\rangle_{\text{nas}} = \frac{1-P_{12}}{\sqrt{2}}
   |\bm{k}\,S S_z\, T T_z\rangle = \sum_{lm} 
  \frac{1-(-1)^{l+S+T}}{\sqrt{2}} Y^*_{lm}(\hat{\bm{k}}) |k\,lm\,S S_z\, T
  T_z\rangle = \sum_{lm}
  Y^*_{lm}(\hat{\bm{k}}) |k\,lm\,S S_z\, T
  T_z \rangle_{\text{nas}}. 
\end{equation}

The calculation of the matrix element for NN bremsstrahlung requires
the evaluation of the spatial trace of the hadronic
tensor~\citep{Raffelt.Seckel:1995,Hannestad.Raffelt:1998}. This
includes contributions from the eight diagrams given in Figure \ref{fig:NNB}~\citep[see also][]{Friman.Maxwell:1979}, which give for the $nn$ or $pp$ channel
\begin{eqnarray}
  \label{eq:8}
  \bar{H}^{(nn)} = \bar{H}^{(pp)} & = & \frac{1}{3} \sum_{s_as_bs_cs_d u}|{}_{\text{nas}}
  \langle \bm{k}_f\,s_cs_d | \bm{\mathcal{O}}_u |
 \bm{k}_i \,s_as_b \rangle_{\text{nas}}|^2 
 \nonumber \\
  & = &  \frac{1}{3\cdot4} 
  \sum_{s_as_bs_cs_d u} (-1)^{1+u} \langle \bm{k}_f\,
  s_cs_d|\bm{\mathcal{O}}_u(1-P_{12})|\bm{k}_i\,s_as_b\rangle \langle
  \bm{k}_i\,s_as_b|\tilde{\bm{\mathcal{O}}}_{-u}(1-P_{12})|\bm{k}_f\,
  s_cs_d\rangle, 
\end{eqnarray}
where the sum on $u$ runs over the spherical components, $0$ and
$\pm1$, of the vector operator $\bm{\mathcal{O}}$ that satisfies
$(\bm{\mathrm{O}}_u)^\dagger = (-1)^{1+u}
\tilde{\bm{\mathrm{O}}}_{-u}$.  We have used non-antisymmetric states
to explicitly show the direct and exchange contributions. Expressions
for the operator $\bm{\mathcal{O}}$ within the long-wavelength limit
used in the present work will be provided later. For the moment, we
keep the formalism fully general. The obtained expressions are then
also applicable for more sophisticated treatments of the nuclear weak
current and/or the intermediate nucleon propagator. Using
states of total spin $S$ and isospin $T$ we have

\begin{subequations}
\begin{eqnarray}
  \label{eq:11}
  \bar{H}^{(nn)} & = & \frac{-1}{3\cdot4} \sum_{\substack{S_f S_{z,f}\\
    S_i S_{z,i}\\u}} (-1)^{u} \langle \bm{k}_f\,S_f S_{z,f}\, 1 \, 1
                     |\bm{\mathcal{O}}_u(1-P_{12})|\bm{k}_i\, S_i
                     S_{z,i}\, 1\, 1\rangle \langle \bm{k}_i\, S_i S_{z,i}\, 1\, 1|\tilde{\bm{\mathcal{O}}}_{-u}(1-P_{12})|\bm{k}_f \, S_f S_{z,f}\, 1\, 1\rangle, \\
  \bar{H}^{(pp)} & = & \frac{-1}{3\cdot4} \sum_{\substack{S_f S_{z,f}\\
    S_i S_{z,i}\\u}} (-1)^{u} \langle \bm{k}_f \, S_f S_{z,f}\, 1 \, {-1}
                     |\bm{\mathcal{O}}_u(1-P_{12})|\bm{k}_i\, S_i
                     S_{z,i}\, 1\, {-1}\rangle \langle \bm{k}_i\, S_i
                     S_{z,i}\, 1\,
                     {-1}|\tilde{\bm{\mathcal{O}}}_{-u}(1-P_{12})|\bm{k}_f
                     \, S_f S_{z,f}\, 1\, {-1}\rangle. \nonumber \\
\end{eqnarray}
\end{subequations}

For the $np$ case the direct and exchange terms
correspond to physically different processes whose contribution needs
to be summed. We obtain

\begin{equation}
  \label{eq:12}
  \bar{H}^{(np)} = \frac{1}{3} \sum_{\substack{S_f S_{z,f}\\
    S_i S_{z,i}\\u}} |\langle\bm{k}_f 
  S_f\, S_{z,f}|\bm{O}_u(1-P_{12})|\bm{k}_i S_i\, S_{z,i}\rangle|^2,
\end{equation}
which gives for isospin states
\begin{equation}
  \label{eq:13}
 \bar{H}^{(np)} = \frac{-1}{3\cdot 4} \sum_{\substack{S_f S_{z,f} T_f\\
    S_i S_{z,i} T_i \\ T'_f T'_i u}} (-1)^{u} \langle\bm{k}_f \, 
  S_f S_{z,f}\,T_f 0|\bm{\mathcal{O}}_u(1-P_{12})|\bm{k}_i\, S_i S_{z,i}\, T_i
  0\rangle \langle \bm{k}_i\, S_i S_{z,i}\, T'_i
  0|\tilde{\bm{\mathcal{O}}}_{-u}(1-P_{12})|\bm{k}_f \, 
  S_f S_{z,f}\,T'_f 0\rangle. 
\end{equation}

In the following we provide formulas to evaluate the necessary matrix
elements using standard angular momentum algebra. We need to evaluate
the following matrix elements:

\begin{equation}
  \label{eq:14}
 \bar{H}(T_z) = \frac{-1}{3\cdot 4} \sum_{\substack{S_f S_{z,f} T_f\\
    S_i S_{z,i} T_i \\ T'_f T'_i u}} (-1)^{u} \langle\bm{k}_f \, 
  S_f S_{z,f}\,T_f T_z|\bm{\mathcal{O}}_u(1-P_{12})|\bm{k}_i\, S_i S_{z,i}\, T_i
  T_z\rangle \langle \bm{k}_i\, S_i S_{z,i}\, T'_i
  T_z|\tilde{\bm{\mathcal{O}}}_{-u}(1-P_{12})|\bm{k}_f \, 
  S_f S_{z,f}\,T'_f T_z\rangle,   
\end{equation}
with $\bar H^{(nn)}=\bar H(T_z=1)=\bar H^{(pp)}=\bar H(T_z=-1)$ and
$\bar H^{(np)}=\bar H^{(pn)}=\bar H(T_z=0)$. 

Following~\citet{Bartl.Pethick.Schwenk:2014,Bartl:2016}, we proceed by
doing a partial wave expansion using Equation~\eqref{eq:pw},
introducing states of total angular momentum using Equation~\eqref{eq:coupleJ}, and then
writing the product of spherical harmonics with the same arguments as
a sum over spherical harmonics. In addition, we use the Wigner-Eckart
theorem to explicitly perform the sum over projections. This gives

\begin{equation}
\label{eq:Hallm}
\begin{aligned}
\bar{H}(T_z)  =& \frac{1}{3\cdot 4} \sum_{\substack{J_i J_f J'_i J'_f\\ T_i T_f T'_i T'_f \\l_i l_f
  l'_i l'_f \\ S_i S_f L L'}}\sum_{\substack{M_i M_f M'_i M'_f\\ m_im_fm'_im'_f\\S_{z,i}S_{z,f}MM'u}}  
 (-1)^{J_f-M_f+J'_i-M'_i+u+1+l_f-S_f+ M_f+l'_f-S'_f+
   M'_f+l_i-S_i+M_i+l'_i-S'_i+M'_i}  \\
 & 
 \left(\begin{array}{ccc}
         J_f & 1 & J_i \\
         -M_f & u & M_i
 \end{array}\right)  
\left(\begin{array}{ccc}
        J'_i & 1 & J'_f \\
        -M'_i & -u & M'_f
      \end{array}\right)
    \left(\begin{array}{ccc}
        l_i & S_i & J_i \\
        m_i & S_{z,i} & -M_i
     \end{array}\right) 
   \left(\begin{array}{ccc}
       l_f & S_f & J_f \\
       m_f & S_{z,f} & -M_f
    \end{array}\right) 
  \left(\begin{array}{ccc}
      l'_i & S_i & J'_i \\
      m'_i & S_{z,i} & -M'_i
    \end{array}\right) \\
&  \left(\begin{array}{ccc}
     l'_f & S_f & J'_f \\
     m'_f & S_{z,f} & -M'_f
    \end{array}\right)
    \left(\begin{array}{ccc}
      l_i &l'_i &L \\
            -m_i &m'_i &M
    \end{array}\right)
  \left(\begin{array}{ccc}
          l_i &l'_i &L \\
          0 &0&0
  \end{array}\right)  
\left(\begin{array}{ccc}
     l_f &l'_f &L' \\
      m_f &-m'_f &M'
      \end{array}\right)
    \left(\begin{array}{ccc}
      l_f &l'_f &L' \\
      0 &0&0
     \end{array}\right)\\
& (-1)^{m_i+m'_f+M+M'} \frac{[J_i J_f J'_i J'_f l_i l_f l'_i l'_f L L']}{4\pi}
Y_{L-M}(\hat{\bm{k}}_i)Y_{L'-M'}(\hat{\bm{k}}_f)\\
& \langle k_f \, l_f 
  S_f J_f\,T_f T_z||\bm{\mathcal{O}}(1-P_{12})||k_i \, l_i S_i J_i\, T_i
  T_z\rangle \langle k_i\, l'_i S_i J'_i\, T'_i
  T_z||\tilde{\bm{\mathcal{O}}}(1-P_{12})||k_f \, 
  l'_f S_f J'_f \,T'_f T_z\rangle,
\end{aligned}   
\end{equation}
where the matrix elements are reduced in total angular momentum space
but not in isospin and we have introduced the notation
$[J]\equiv\sqrt{2J+1}$. The sums over projection quantum numbers can
now be performed using the following relations between 3-$j$ and 6-$j$
symbols, the orthogonality of 3-$j$ symbols, and the spherical
harmonics addition theorem:

\begin{subequations}
  \begin{equation}
    \label{eq:15}
    \begin{aligned}
\sum_{m_i m'_i S_{z,i}} & (-1)^{l_i+l'_i+S_i+m'_i+m_i-S_{z,i}}
\left(\begin{array}{ccc} J_i & l_i & S_i \\ -M_i & m_i & S_{z,i} \end{array}\right) 
\left(\begin{array}{ccc} l'_i & J'_i & S_i \\ -m'_i & M'_i & -S_{z,i} \end{array}\right) 
\left(\begin{array}{ccc} l'_i & l_i & L \\ m'_i & -m_i & M \end{array}\right) \\
& = \left(\begin{array}{ccc} J_i & J'_i & L \\ -M_i & M'_i & M \end{array}\right)  
\left\{\begin{array}{ccc} J_i & J'_i & L \\ l'_i & l_i & S_i \end{array}\right\},      
    \end{aligned}
\end{equation}
\begin{equation}
  \label{eq:16}
    \begin{aligned}
\sum_{m_f m'_f S_{z,f}}  & (-1)^{l_f+l'_f+S_f+m'_f+m_f-S_{z,f}}
\left(\begin{array}{ccc} J_f & l_f & S_f \\ -M_f & m_f & S_{z,f} \end{array}\right) 
\left(\begin{array}{ccc} l'_f & J'_f & S_f \\ -m'_f & M'_f & -S_{z,f} \end{array}\right) 
\left(\begin{array}{ccc} l'_f & l_f & L' \\ m'_f & -m_f & -M' \end{array}\right) \\
& = \left(\begin{array}{ccc} J_f & J'_f & L' \\ -M_f & M'_f & -M' \end{array}\right) 
\left\{\begin{array}{ccc} J_f & J'_f & L' \\ l'_f & l_f & S_f \end{array}\right\},  
\end{aligned}
\end{equation}
\begin{equation}
  \label{eq:17}
    \begin{aligned}
 \sum_{M_i M'_i u} & (-1)^{J'_i+J_i+1+M'_i+M_i-u}     
\left(\begin{array}{ccc} J_f & J_i & 1 \\ -M_f & M_i & u \end{array}\right) 
\left(\begin{array}{ccc} J'_i & J'_f & 1 \\ -M'_i & M'_f & -u \end{array}\right) 
\left(\begin{array}{ccc} J'_i & J_i & L \\ M'_i & -M_i & M \end{array}\right) \\
& = \left(\begin{array}{ccc} J_f & J'_f & L \\ -M_f & M'_f & M \end{array}\right) 
\left\{\begin{array}{ccc} J_f & J'_f & L \\ J'_i & J_i & 1 \end{array}\right\}, 
\end{aligned}
\end{equation}
\begin{equation}
  \label{eq:18}
  \sum_{M_f M'_f} \left(\begin{array}{ccc} J_f & J'_f & L \\ -M_f & M'_f & M \end{array}\right) 
\left(\begin{array}{ccc} J_f & J'_f & L' \\ -M_f & M'_f & -M' \end{array}\right) = \frac{1}{2L+1} \delta_{LL'}\delta_{M-M'}, 
\end{equation}
\begin{equation}
  \label{eq:19}
\sum_{MM'} (-1)^M Y_{L-M}(\hat{\bm{k}}_i) Y_{L-M'}(\hat{\bm{k}}_f)
\delta_{M-M'}= \sum_M Y^*_{LM}(\hat{\bm{k}}_i) Y_{LM}(\hat{\bm{k}}_f) =
  P_L(\hat{\bm{k}}_i \cdot \hat{\bm{k}}_f) \frac{2L+1}{4\pi},   
\end{equation}
\end{subequations}
to finally obtain

\begin{equation}
\label{eq:Hsum}
\begin{aligned}
\bar{H}(T_z)  =&\frac{1}{3\cdot 4(4\pi)^2} \sum_{\substack{J_i J_f
    J'_i J'_f \\ T_i T_f
  T'_i T'_f \\ l_i l_f  l'_i l'_f \\S_i S_f L}} 
 (-1)^{S_f+S_i+J_i+L+J'_f} [J_i J_f J'_i J'_f l_i
 l_f l'_i l'_f] [L]^2 \\
 & \left\{\begin{array}{ccc} J_i & J'_i & L \\ l'_i & l_i &
                                                            S_i \end{array}\right\} \left\{\begin{array}{ccc} J_f & J'_f & L \\ l'_f & l_f & S_f \end{array}\right\}
\left\{\begin{array}{ccc} J_f & J'_f & L \\ J'_i & J_i & 1 \end{array}\right\}
\left(\begin{array}{ccc} l_i &l'_i &L \\ 0 &0&0\end{array}\right)  
\left(\begin{array}{ccc} l_f &l'_f &L \\ 0 &0&0\end{array}\right)
P_L(\hat{\bm{k}}_i \cdot \hat{\bm{k}}_f) \\
& \langle k_f \, l_f 
  S_f J_f\,T_f T_z||\bm{\mathcal{O}}(1-P_{12})||k_i \, l_i S_i J_i\, T_i
  T_z\rangle \langle k_i\, l'_i S_i J'_i\, T'_i
  T_z||\tilde{\bm{\mathcal{O}}}(1-P_{12})||k_f \, 
  l'_f S_f J'_f \,T'_f T_z\rangle.
\end{aligned}
\end{equation}

Equation~\eqref{eq:Hsum} is valid for any vector (rank 1) operator and
hence can be used even in calculations that consider the weak hadronic
current beyond the leading-order approximation, and it allows for the inclusion of
two-body currents. At leading order in the weak current and within the
long-wavelength limit the operator $\bm{\mathcal{O}}$ can be expressed
as

\begin{equation}
  \label{eq:9}
  \bm{\mathcal{O}}_u = \frac{1}{\omega} [\mathcal{T},
  \bm{\mathcal{Y}}_u]', \quad \tilde{\bm{\mathcal{O}}}_u = \frac{1}{\omega} [\mathcal{T}^\dagger,
  \bm{\mathcal{Y}}_u]', \quad
  \text{with}\quad \bm{\mathcal{Y}}_u = \sum_r \bm{\sigma}_u^{(r)} \bm{\tau}^{(r)}_z,
\end{equation}
where $\mathcal{T}$ is the $T$-matrix and the sum in $r$ runs over the
two initial or final nucleons. We have introduced the isospin operator
$\bm{\tau}_z |n\rangle = |n\rangle, \bm{\tau}_z |p\rangle =
-|p\rangle$ to make clear the spin-isospin dependence of the
operator. The factor $1/\omega$ originates from the non-relativistic
propagator of the nucleon to which the weak interaction is attached.
The prime in the commutator denotes 
  that the $T$-matrix is evaluated at different values of the energy
  for the first (``positive'') and second (``negative'') terms:
\begin{equation} 
\label{eq:comm}   
\langle \bm{k}_f| [\mathcal{T}, A]' |\bm{k}_i\rangle \equiv \langle \bm{k}_f|\mathcal{T}(E_{k_f}) A - A \mathcal{T}(E_{k_i}) |\bm{k}_i\rangle,
\end{equation} 
with $A$ an arbitrary operator and $E_k = k^2/m_N$. Finally, for
the reduced matrix elements of the operators $\bm{\mathcal{O}}$ and
$\tilde{\bm{\mathcal{O}}}$ we have

\begin{equation}
  \label{eq:20}
  \begin{aligned}
 \frac{1}{2} \langle k_f \, l_f 
  S_f J_f\,T_f T_z||\bm{\mathcal{O}}(1-P_{12})||k_i \, l_i S_i J_i\, T_i
  T_z\rangle = \frac{1}{\omega}\Bigl\{ & \mathcal{T}_{l_fl_i}^{S_f J_f
    T_f}(k_f,k_i; E_{k_f}) \langle k_i \, l_i S_f J_f\, T_f
    T_z||\bm{\mathcal{Y}}||k_i\, l_i S_i J_i\, T_i T_z \rangle\\
     & -
    \langle k_f\, l_f S_f J_f\, T_f T_z|| \bm{\mathcal{Y}} ||k_f\, l_f
    S_i J_i\, T_i T_z \rangle \mathcal{T}_{l_fl_i}^{S_i J_i
      T_i}(k_f,k_i;E_{k_i})\Bigr\},      
  \end{aligned}
\end{equation}

\begin{equation}
  \label{eq:21}
  \begin{aligned}
  \frac{1}{2}\langle k_i\, l'_i S_i J'_i\, T'_i
  T_z||\tilde{\bm{\mathcal{O}}}(1-P_{12})||k_f \, 
  l'_f S_f J'_f \,T'_f T_z\rangle = \frac{1}{\omega}\Bigl\{ &
  \bigl[\mathcal{T}_{l'_fl'_i}^{S_i J'_i T'_i}(k_f,k_i;E_{k_i})\bigr]^*
  \langle k_f\, l'_f S_i J'_i\, T'_i T_z||\bm{\mathcal{Y}}||k_f\, l'_f
  S_f J'_f\, T'_f T_z \rangle\\ 
     & -
 \langle k_i\, l'_i S_i J'_i\, T'_i T_z||\bm{\mathcal{Y}}||k_i\, l'_i S_f J'_f\,
 T'_f T_z \rangle \bigl[\mathcal{T}_{l'_fl'_i}^{S_f J'_f
   T'_f}(k_f,k_i;E_{k_f})\bigr]^* \Bigr\}, 
  \end{aligned}
\end{equation}
where we have introduced the shorthand notation for the vacuum $T$-matrix elements

\begin{equation}
  \label{eq:22}
  \begin{aligned}
  \mathcal{T}_{l_fl_i}^{S J T}(k_f,k_i;E) \equiv &\, {}_{\text{nas}}\langle k_f \, l_f 
  S J M\,T T_z|\mathcal{T}(E)|k_i \, l_i S J M \, T
  T_z\rangle_{\text{nas}} \\
  = & \frac{1}{2}
  \langle k_f \, l_f 
  S J M\,T T_z|\mathcal{T}(E)(1-P_{12})|k_i \, l_i S J M \, T
  T_z\rangle,    
  \end{aligned}
\end{equation}
with $E$ given by $E_{k_i}= k_i^2/m_N$ or $E_{k_f}= k_f^2/m_N$ for
the initial or the final nucleon pair to be on-shell. It can be easily generalised to
the case of using the in-medium $T$-matrix, where one needs to replace all the
$T$-matrix elements, $\mathcal{T}_{l_fl_i}^{S J T}(k_f,k_i;E_{k})$,
by $\mathcal{T}_{l_fl_i}^{S J T}(k_f,k_i;K,\Omega_{k})$, with
$\Omega_{k}=K^2/(4m_N)+k^2/m_N$ and $k$ being either $k_i$ or $k_f$.     
The reduced matrix elements of the operator $\bm{\mathcal{Y}}$
are

\begin{equation}
  \label{eq:23}
  \begin{aligned}
    \langle k \, l S_f J_f\, T_f T_z||\bm{\mathcal{Y}}||k\, l S_i J_i\,
    T_i T_z \rangle = & 6(-1)^{T_f-T_z+l+S_i+J_f+1} [S_i S_f J_i J_f T_i
    T_f] \Bigl[(-1)^{S_f+T_f}+(-1)^{S_i+T_i}\Bigr]\\
        & \left(\begin{array}{ccc}
            T_f & 1 & T_i \\
            -T_z & 0 & T_z
          \end{array}\right)
          \left\{\begin{array}{ccc}
          S_f & S_i & 1\\
          J_i & J_f & l
          \end{array}\right\}
        \left\{\begin{array}{ccc}
          \frac{1}{2} & \frac{1}{2} & 1\\
          S_i & S_f & \frac{1}{2}\\
        \end{array}\right\}               
        \left\{\begin{array}{ccc}
          \frac{1}{2} & \frac{1}{2} & 1\\
          T_i & T_f & \frac{1}{2}\\
        \end{array}\right\}.               
  \end{aligned}
\end{equation}

Notice that, due to the fact that we use normalized antisymmetric
states, the sums in Equation~\eqref{eq:Hsum} are restricted to
combinations of $l$, $S$, and $T$ such as $l+S+T$ is odd; see
Equation~\eqref{eq:7}. For $T_z=1$ we have $S_i=S_f=1$, while for $T_z=0$ we
have $T_i=0, T_f=1$ or $T_i=1, T_f=0$ and either $S_f>0$ or
$S_i>0$. 

\section{Numerical table for our new structure function based on the vacuum $T$-matrix}
\label{sec:table}    

We provide a numerical table of the normalized structure function
$S_\sigma(\omega)$ based on the vacuum $T$-matrix at
\url{http://github.com/dcpresn23/Tables-for-bremsstrahlung-Rate-in-SN}
(`S\_Table\_Tv.dat'), where the multiple-scattering effects are
included but the RPA correlation is not considered. The table covers a
wide range of conditions relevant to SN matter with $2~\rm{MeV} \le T\le 50$
MeV (25 bins), $10^{-4}~\rm{fm^{-3}}\le n_B \le 1$ fm$^{-3}$ (37 bins), and
$0 \le Y_e \le 0.5$ (26 bins). The structure function is evaluated at
$\omega_i=10^{0.1 \times i-1.4}$ with $i=1,2,...,40$. The maximal
energy transfer is $\omega_\mathrm{max}\simeq 400$ MeV. Note that our
results may be inaccurate at densities higher than the saturation
density, since the medium effects and three-body force can be
important. However, we expect that neutrinos are
trapped in such conditions. To use the table in SN simulations, one needs to do 4D
interpolations over $T, n_B, Y_e$, and $\omega$ to obtain $S_\sigma(\omega)$
in each condition. Since we use the same notation, the new structure
function can be implemented in a similar way to the fitting formula
from \cite{Hannestad.Raffelt:1998}.
%\end{thebibliography}

%% This command is needed to show the entire author+affilation list when
%% the collaboration and author truncation commands are used.  It has to
%% go at the end of the manuscript.
%\allauthors

%% Include this line if you are using the \added, \replaced, \deleted
%% commands to see a summary list of all changes at the end of the article.
%\listofchanges

%\bibliographystyle{aasjournal}
%\bibliography{../biblio/bibliography}
\bibliography{refs}

\listofchanges 
\end{document}